 \documentclass[aps,prd,superscriptaddress,showpacs,twocolumn,preprintnumbers]{revtex4}

\usepackage{epsfig,dcolumn}
\usepackage{graphicx}
\usepackage{color}

\usepackage{amsmath}
\usepackage{amsfonts}
\usepackage{amssymb}
\usepackage{graphicx}
\usepackage{type1cm}\usepackage{eso-pic}

\newcommand{\be}{\begin{equation}}
\newcommand{\ee}{\end{equation}}

\newcommand{\im}{\mathrm{Im}\,}
\newcommand{\re}{\mathrm{Re}\,}

\newcommand{\degree}{{\rm o}}
\newcommand{\gev}{{\rm GeV}}

\usepackage{bm} 

\begin{document}

\title{$\pi \pi \rightarrow K \bar{K}$ scattering up to 1.47 GeV with hyperbolic dispersion relations.} 

\author{J.R.~Pelaez}
\affiliation{Departamento de F\'isica Te\'orica, Universidad Complutense de Madrid, 28040 Madrid, Spain}
\author{A.Rodas}
\affiliation{Departamento de F\'isica Te\'orica, Universidad Complutense de Madrid, 28040 Madrid, Spain}

\begin{abstract}
In this work we provide a dispersive analysis of $\pi\pi \rightarrow K\bar{K}$ scattering.
For this purpose we present a set of partial-wave hyperbolic dispersion relations using a family of hyperbolas that maximizes the applicability range of the hyperbolic dispersive representation, which we have extended up to 1.47 GeV. We then use these equations first to test simple fits to different and often conflicting data sets, also showing that some of these data and some popular parameterizations of these waves fail to satisfy the dispersive analysis.
Our main result is obtained after imposing these new relations as constraints on the data fits. We thus provide simple and precise parameterizations for the S, P and D waves that describe the experimental data from $K\bar K$ threshold up to 2 GeV, while being consistent with crossing symmetric partial-wave dispersion relations up to their maximum applicability range of 1.47 GeV. For the $S$-wave we have found that two solutions describing two conflicting data sets are possible. The dispersion relations also provide a representation for $S$, $P$ and $D$ waves in the pseudo-physical region. 
\end{abstract}
\maketitle

\section{Introduction}

The scattering of pions and kaons is interesting for several reasons:
First, by itself, in order to test and understand the dynamics of these particles, which are
the pseudo-Goldstone Bosons of the QCD spontaneous chiral symmetry breaking. 
Second, because these scattering processes
are one of the main sources of information on the existence and parameters 
of several meson resonances. In particular, this is the case of light scalar mesons, 
whose very existence, nature and classification are still a matter of debate
(see the note on light scalars in the Review of Particle Properties (RPP)\cite{RPP}).
These resonances are very relevant
for the identification of glueballs, tetraquaks or molecular states that lie
beyond the ordinary meson states of the naive quark model. 
Finally, being the lightest mesons, final state interactions (FSI) of 
pions and kaons play an essential
role in the description of many hadronic processes. 
The unprecedented statistical samples 
obtained in the last years on different hadronic experiments
and the even more ambitious plans for future facilities
have provoked a renovated interest for
precise and rigorous analyses of existing meson-meson scattering data, superseding simple model descriptions.

Unfortunately, most of the data on meson-meson scattering 
\cite{Protopopescu:1973sh,Grayer:1974cr,Estabrooks:1974vu,datapiK,Cohen:1980cq,Etkin:1981sg,Longacre:1986fh,Lindenbaum:1991tq}
are extracted indirectly
from meson-nucleon to meson-meson-nucleon reactions. This extraction is complicated, relying 
on some model assumptions, and for this reason 
it is affected with large systematic uncertainties, which can be estimated from the
differences between data sets from different experiments (and for $\pi\pi$ scattering
even within data sets from the same experiment \cite{Protopopescu:1973sh,Grayer:1974cr}). Moreover, the description
of these data is frequently done in terms of meson-meson models which can lead to artifacts and 
unreliable determinations of resonances and their parameters.
It is for these reasons that dispersive techniques are required. 

Dispersion relations are 
the mathematical expression of causality and crossing. They relate the amplitude at a given energy to
integrals of the amplitude and can be used as consistency 
tests of the experimental data or as constraints on the fits. 
We will make both uses here.
For dispersive integrals to be evaluated just over the physical region,
crossing must be used and two main kinds of dispersion relations appear then: Forward Dispersion Relations (FDRs) and those for partial waves generically know as Roy or Roy-Steiner equations \cite{Roy:1971tc,Steiner:1971ms}, depending on whether the scattering occurs among particles with equal or different masses.  
FDRs are rather simple and easily
extended to arbitrary energies. They have been recently applied to constrain
$\pi\pi$ \cite{Pelaez:2004vs,Kaminski:2006yv,Kaminski:2006qe,GarciaMartin:2011cn} and $K\pi$ \cite{Pelaez:2016tgi} 
scattering amplitudes that will be used as input
in some stages of the present work. 
Roy-like equations are a complicated system of coupled equations,
limited in practice to energies of $\mathcal{O}(1\, \gev)$ for meson-meson scattering.
However, they provide a rigorous continuation to the complex plane 
that allows for a precise and model independent determination of resonances.
Actually, it was only in 2012 that the RPP\cite{RPP} considered settled the issue of
the existence and parameters of the much debated scalar $f_0(500)$ resonance \cite{Pelaez:2015qba}, 
traditionally known as $\sigma$-meson, 
and to a very large extent this was due to the results of dispersive analyses
of $\pi\pi$ scattering amplitudes with versions of Roy equations \cite{Ananthanarayan:2000ht,sigmaroy}.
Similarly, the scalar $K_0^*(800)$ or $\kappa$-meson has also been obtained from $\pi K$ scattering 
using dispersive methods \cite{kappauchpt,kappadispersive}, 
the most reliable value \cite{DescotesGenon:2006uk} being the Roy-Steiner method based on hyperbolic dispersion relations \cite{Buettiker:2003pp}, but
according to the RPP this resonance still ``needs confirmation'' \cite{RPP}.
Roy-Steiner equations have also been applied recently to $\pi N$ scattering \cite{Ditsche:2012fv}
and for $\gamma\gamma\rightarrow\pi\pi$ \cite{Hoferichter:2011wk}.
For meson-resonances beyond $\sim$1 GeV, Roy-like equations are not used in practice, but other
analytic tools have been recently applied \cite{Pades} to extract resonance poles from the 
description of amplitudes in the physical region constrained with dispersion relations, thus 
minimizing the model-dependence.

The purpose of this paper is to obtain a set of 
simple $\pi\pi  \rightarrow K\bar{K}$ scattering parameterizations 
satisfying Roy-Steiner dispersion relations that can be easily used later on
both by theoreticians and experimentalists, 
as has already been the case of previous works for $\pi\pi$ and $\pi K$ scattering.
The motivations to study $\pi\pi \rightarrow K\bar{K}$ are the ones
explained above for meson-meson scattering in general: i) a rigorous
$\pi\pi\rightarrow K\bar{K}$ description is a necessary input for further studies of resonances (like scalars in the 1 to 1.6 GeV range), in particular in order to compare their $\pi\pi$ and $K\bar{K}$ couplings, ii) it is also an essential ingredient in the Roy-Steiner study of 
$K\pi$ scattering and the determination of the controversial $K^*_0(800)$-meson
(whose determination is one of the goals of a recent proposal at JLab \cite{Amaryan:2017ldw})
iii) the $\pi\pi  \rightarrow K\bar{K}$ amplitude also influences, via unitarity, the $\pi\pi  \rightarrow \pi\pi$ and $\pi\pi\rightarrow NN$
amplitudes, and consequently those of $KN$ and $\bar{K} N$ scattering.
Finally $\pi\pi  \rightarrow K\bar{K}$
is a very relevant ingredient in the FSI of numerous hadron decays. For instance,
the role of $\pi\pi\rightarrow K\bar{K}$ re-scattering has gained a renewed interest due to the
recent observation of a large CP violation in recent studies at LHCb \cite{CPV}, although
the amplitude used for such studies has been approximated with simple models and 
the amplitudes obtained here could be used to avoid such assumptions in further studies which are under way. Finally, lattice calculations of the coupled channel $\pi\pi$, $K\bar{K}$, $\eta\eta$ scattering
have appeared very recently \cite{Briceno:2016mjc}. Although these calculations
are performed still at relatively high pion masses, the physical point where one can compare with 
our actual $\pi\pi\rightarrow K\bar{K}$ parameterizations could be accessible soon.

Dispersive studies of $\pi\pi\rightarrow K\bar{K}$ scattering and its relation to 
$\pi K\rightarrow \pi K$ scattering were first performed in the seventies
\cite{Nielsen:1973au,Johannesson:1974ma,Palou:1975uu,Johannesson:1976qp}. 
It was soon clear that the formalism of fixed-t dispersion relations combined with hyperbolic dispersion relations (HDR) for partial waves \cite{Steiner:1971ms} was best suited to study the physical regions of both channels simultaneously \cite{Johannesson:1974ma,Johannesson:1976qp}. However, 
$\pi\pi\rightarrow K\bar{K}$ data was scarce and these analyses only allowed for
crude checks of low-energy scalar partial waves, frequently focusing on threshold parameters
and the non-physical region between the two-pion and the two-kaon thresholds (or at most up to 1100 MeV). For a review of the theoretical and experimental situation until 1978 we refer to \cite{Lang:1978fk}.

The main experimental results on $\pi\pi\rightarrow K\bar{K}$ partial waves, that will be thoroughly analyzed in this work, 
were obtained in the early eighties  \cite{Cohen:1980cq,Etkin:1981sg}, indirectly
from $\pi N\rightarrow K\bar{K} N'$ reactions. They extend from energies 
very close to the $K\bar{K}$ threshold up to 1.6 GeV.
Several models exist in the literature describing
these $\pi\pi\rightarrow K\bar{K}$ data \cite{models}, in particular 
with unitarized chiral Lagrangians \cite{kappauchpt,chiralmodels}. These works are of relevance for studies of  $f_0$ resonances and glueballs in that range.

A renewed interest on dispersive analysis of $\pi\pi\rightarrow K\bar{K}$ at the turn of the century
was triggered by the need for precise determinations of threshold parameters and Chiral Perturbation Theory 
low energy constants. Actually, sum rules for $\pi K$ were obtained from a Roy-Steiner type of equations from HDR \cite{Ananthanarayan:2001uy} in which the $\pi\pi\rightarrow K\bar{K}$ amplitude in the unphysical
 region was obtained as a solution of a dispersive Mushkelishvili-Omn\'es problem. 
The $\pi\pi\rightarrow K\bar{K}$ partial-wave data of \cite{Cohen:1980cq,Etkin:1981sg} was used as input.
However, no dispersive analysis of these data has been carried out beyond the  $K\bar K$ threshold, mostly due to the relatively low applicability limit of the HDR along the $su=b$ hyperbolas used in those works.
It was nevertheless shown that an extrapolation of the HDR solutions beyond their applicability region
was fairly close to the data. Finally, in \cite{Buettiker:2003pp} a Roy-Steiner type of analysis
was performed to obtain solutions for the $\pi K$ elastic amplitudes, using once again as input 
the $\pi\pi\rightarrow K\bar{K}$ amplitudes in the physical region. This study was the basis for 
confirming the existence of the $K_0^*(800)$ meson through a dispersive analysis \cite{DescotesGenon:2006uk}.

The aim of this work is then to provide a simple set of $\pi\pi\rightarrow K\bar{K}$
parameterizations that describe the data up to 2 GeV while also satisfying dispersive constraints 
in the whole region from $\pi\pi$ threshold up to 1.47 GeV. 
To this end, we will derive a new set of hyperbolic dispersion relations, 
along $(s-a)(u-a)=b$ hyperbolas, choosing the $a$ parameter to maximize the applicability range which allows us to 
use them up to 1.47 GeV. This will also allow us to test different and often conflicting data sets and popular parameterizations.

The plan of the work is as follows: in Sec.\ref{sec:notation} we will introduce the notation, in Sec.\ref{sec:UFD} we will
present simple unconstrained 
fits to the different $\pi\pi\rightarrow K\bar{K}$ data as well as a Regge formalism
for the high energy part, taking particular care on the determination of uncertainties. In Sec.\ref{sec:HDR} we will derive our new set of HDR, i.e. Roy-Steiner like equations for partial waves,
and formulate the Mushkelishvili-Omn\'es problem used for both the unphysical region below $K \bar K$ threshold and the physical region up to 1.47 GeV. 
In Sec.\ref{sec:Consistency} we will first use these equations as checks
for the unconstrained parameterizations. Finally, in Sec.\ref{sec:CFD} we will impose the 
new relations on the data fits. This will lead to the desired constrained fits to data 
satisfying the analyticity requirements, which are the main results of this work. 
In Sec.\ref{sec:Conclusions} we will summarize our findings and conclude.

\section{Kinematics and Notation}
\label{sec:notation}

Throughout this work we will be working in the isospin limit
of equal mass for all pions,
$m_\pi=139.57\,$MeV, and equal mass for all kaons, $m_K=496\,$MeV. 

Crossing symmetry relates the $\pi\pi\rightarrow K\bar{K}$ amplitudes
to those of $\pi K$ scattering. It is then customary to use the standard Mandelstam variables $s,t,u$
for $\pi K$ scattering, satisfying $s+t+u=2(m_\pi^2+m_K^2)$ and write
\begin{eqnarray}
G^0(t,s,u)&=&\sqrt{6}F^+(s,t,u),\nonumber\\
G^1(t,s,u)&=&2F^-(s,t,u),
\end{eqnarray}
where $G^I$ are the fixed isospin $I=0,1$ amplitudes of $\pi\pi\rightarrow K\bar{K}$
whereas the $F^{\pm}$ are the $s\leftrightarrow u$ symmetric and antisymmetric $\pi K$ amplitudes, respectively. The latter are
defined as
\begin{eqnarray}
F^+(s,t,u)&=&\frac{1}{3}F^{1/2}(s,t,u)+\frac{2}{3}F^{3/2}(s,t,u),\nonumber\\
F^-(s,t,u)&=&\frac{1}{3}F^{1/2}(s,t,u)-\frac{1}{3}F^{3/2}(s,t,u),
\end{eqnarray}
where now $F^{I}$ are the fixed isospin $I=1/2,3/2$ amplitudes of $\pi K$ scattering.
These satisfy:
\begin{equation}
F^{1/2}(s,t,u)=\frac{3}{2} F^{3/2}(u,t,s)-\frac{1}{2}F^{3/2}(s,t,u),
\end{equation}
from where the $s\leftrightarrow u$ symmetry properties of $F^\pm$ follow.

In this work we will also use the partial-wave decompositions of the $\pi K$
and $\pi\pi\rightarrow K\bar{K}$ scattering amplitudes, defined as follows:
\begin{eqnarray}
&&F^I(s,t,u)=16\pi \sum_\ell{(2\ell+1)P_\ell(z_s) f^I_\ell(s)},\label{ec:pwexpansion} \\
&&G^I(t,s,u)=16\pi\sqrt{2} \sum_\ell{(2\ell+1)(q_{\pi}q_{K})^\ell P_\ell(z_t) g^I_\ell(t)},\nonumber
\end{eqnarray}
where $q_{\pi}=q_{\pi\pi}(t),q_K=q_{KK}(t)$
 are the CM momenta of the respective $\pi \pi$ and $K\bar{K}$ states, namely
\begin{equation}
q_{12}(s)=\frac{1}{2\sqrt{s}}\sqrt{(s-(m_1+m_2)^2)(s-(m_1-m_2)^2)}.
\end{equation}
Note the $(q_\pi q_K)^\ell$ factors in the partial waves of the $t$-channels, which
are customarily introduced to ensure good analytic properties for $g_\ell(t)$ 
(see \cite{Frazer:1960zza} in the $\pi\pi\rightarrow N\bar{N}$ context).
The scattering angles in the $s$ and $t$ channels are given by:
\begin{equation}
z_s=\cos\theta_s=1+\frac{2st}{\lambda_s},\quad z_t=\cos\theta_t=\frac{s-u}{4q_{\pi}q_K},
\label{eq:anglet}
\end{equation}
where $\lambda_s=(s-(m_\pi+m_K)^2)(s-(m_K-m_\pi)^2)=4s\,q_{K\pi}^2(s)$.

It is also convenient to define $m_\pm=m_K\pm m_\pi$, $\Sigma_{12}=m_1^2+m_2^2$ and $\Delta_{12}=m_1^2-m_2^2$, as well as $t_\pi=4m_\pi^2$, $t_K=4m_K^2$.
In the rest of this work, and unless  
stated otherwise,  $m_1=m_K$, $m_2=m_\pi$, $\Delta=\Delta_{K\pi}$, $\Sigma=\Sigma_{K\pi}$ and $q=q_{K\pi}(s)$.
For later use we define the $K\pi$ scattering lengths as follows:
\begin{equation}
a_0^I=\frac{2}{m_+}f_0^I(m_+^2)
\end{equation}
and similarly for $a_0^\pm$.

Let us recall that in the case when we have two identical particles in the initial state, as it happens with 
two pions in the isospin limit formalism, we define
\begin{equation}
g^I_\ell(t)=\frac{\sqrt{2}}{32\pi(q_{\pi}q_{K})^\ell}\int_{0}^{1}{dz_tP_\ell(z_t)G^I(t,s)}.
\end{equation}
For later use we also write here the explicit expressions for the 
$\ell=0,1,2$ partial waves:
\begin{eqnarray}
g^0_0(t)&=&\frac{\sqrt{3}}{16\pi}\int_{0}^{1}{dz_t F^+(s,t)},\nonumber\\
g^1_1(t)&=&\frac{\sqrt{2}}{16\pi q_{\pi}q_{K}}\int_{0}^{1}{dz_t z_t F^-(s,t)}, \nonumber \\
g^0_2(t)&=&\frac{\sqrt{3}}{16\pi (q_{\pi}q_{K})^2}\int_{0}^{1}{dz_t \frac{3z^2_t-1}{2}F^+(s,t)}.
\label{proj}
\end{eqnarray}
 Finally, the relation with the S-matrix partial waves, which allows for straightforward
comparison with some experimental works,  is: 
 \begin{eqnarray}
 S^{I}_\ell(s)_{\pi \pi\rightarrow \pi\pi}&=&1+i\frac{4q}{\sqrt{s}}f_\ell^I(s)\theta(s-m_+^2), \\
 S^{I}_\ell(t)_{\pi \pi\rightarrow K \bar K}&=&i \frac{4(q_\pi q_K)^{\ell+1/2}}{\sqrt{t}}g^{I_t}_\ell(t)\theta(t-t_K). \nonumber
 \end{eqnarray}




\section{Unconstrained Fits to Data}
\label{sec:UFD}

\subsection{The Data}
\label{subsec:data}

As we have already emphasized in the introduction we will explicitly choose
very simple parameterizations to fit the data, so that they can be used easily 
later on. In this section we will just describe the data without imposing 
dispersion relations. These will be called Unconstrained Fits to Data (UFD). 
In this way the fits to each wave are independent from each other.
Later on we will impose the dispersion relations as 
constraints and obtain the Constrained
Fits to Data (CFD). This will correlate different waves.

The data we will fit are of four types. First,
we will use data on the phases and modulus of the $g^0_0, g^1_1$
partial waves extracted from  $\pi^- p\rightarrow K^-K^+ n$ and 
$\pi^+ n\rightarrow K^-K^+ p$ at the Argonne National Laboratory  \cite{Cohen:1980cq}
and from $\pi^- p\rightarrow K^0_sK^0_s n$ at the Brookhaven National Laboratory in a series of three works 
\cite{Etkin:1981sg,Longacre:1986fh,Lindenbaum:1991tq}, that we will call Brookhaven-I, Brookhaven-II and Brookhaven-III, respectively. 
Second, for the tensor $g^0_2$ wave, data for its modulus was given in
Brookhaven-II and Brookhaven-III, although as we will see the old experimental parameterizations are not
quite compatible with the present resonance parameters listed in the RPP. 
Third, for higher partial waves, which play a very minor role
in the numerics, we use simple resonance parameterizations with their parameters
as quoted in the RPP. Finally, for the high-energy range above 2 GeV
we rely on recent updates \cite{DescotesGenon:2006uk,GarciaMartin:2011cn,Pelaez:2016tgi},
of Regge parameterizations 
\cite{Pelaez:2003ky} 
based on factorization and the phenomenological observations about Regge trajectories
or the Veneziano model \cite{Veneziano:1968yb}.

\subsection{Partial wave fits from $K \bar K$ threshold to 2 GeV}

We now describe our partial-wave parameterizations in the region from $K\bar K$ 
threshold to 2 GeV. 
For all of them we define a modulus and a phase $t^I_\ell=|t^I_\ell|e^{i\phi^I_\ell}$. We will start with the waves that have less controversy on the data sets and 
that, as we will see later, satisfy best our Roy-Steiner-like equations, leaving for the end the 
most difficult one, which is that with $\ell=0,I=0$.
Note that since in the isospin limit all pions are identical particles, Bose statistics applies and $\ell+I$ must be even.

\subsubsection{$\ell=1,I=1$ partial wave}
\label{sec:UFD11}

\begin{figure}
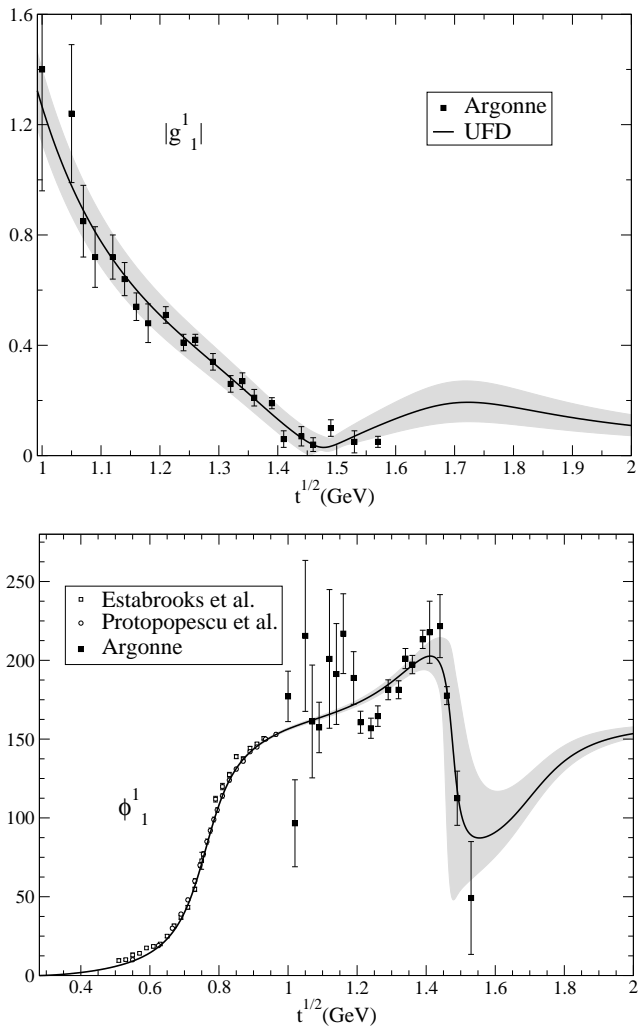

\centering
\includegraphics[scale=0.32]{amplig11.eps}\vspace*{.3cm}
\includegraphics[scale=0.32]{phaseg11.eps}
\caption{\rm \label{fig:g11ufd} 
Modulus and phase of the $g_1^1(t)$ $\pi\pi\rightarrow K \bar K$ partial wave. 
The continuous line 
and the uncertainty band correspond to the UFD 
parameterization described in the text. Note that the phase below $K \bar K$ 
follows that of $I=1,\ell=1$  elastic $\pi\pi$ scattering \cite{GarciaMartin:2011cn}. 
The white circles and squares come from the $\pi\pi$ scattering experiments of
Protopopescu et al. 
\cite{Protopopescu:1973sh} and Estabrooks et al.\cite{Estabrooks:1974vu}, respectively.}
\end{figure}

For the $g_1^1$ partial wave there 
is only data from the Argonne Collaboration 
(Cohen et al. \cite{Cohen:1980cq}), extending 
up to around 1.6 GeV for both the modulus $\vert g_1^1\vert$ and its phase 
$\phi^1_1$. 
Although there is no data on the 1.6 to 2 GeV region, 
which is the starting energy of our Regge parameterizations,
we will see that a rather simple functional form covering the whole range from
$\pi\pi$ threshold up to 2 GeV
satisfies fairly well the Roy-Steiner equations 
 even before imposing them as constraints.
In particular we will use a phenomenological 
parameterization similar to that in \cite{Buettiker:2003pp}:
\begin{eqnarray}
g_1^1(t)&=&\frac{C}{\sqrt{1+r_1 \hat q_\pi^2(t)}\sqrt{1+r_1 \hat q_K^2(t)}} \label{ec:g11} \\
&&\Big\{ \overline{BW}(t)_\rho+(\beta
+\beta_1 \hat q_K^2(t))\overline{BW}(t)_{\rho'}\nonumber\\
&\times&\quad+(\gamma+\gamma_1\hat q_K^2(t))\overline{BW}(t,m)_{\rho''}\Big\}, \nonumber
\end{eqnarray}
where the three vector resonances $\rho(770)$, $\rho'=\rho(1450)$, $\rho''=\rho(1700)$
have been parameterized by a combination of three Breit-Wigner-like shapes:
\begin{eqnarray}
\overline{BW}(t)_V&=&\frac{m_V^2}{m_V^2-t-i\Gamma_V\sqrt{t}\,\frac{2G_\pi(t)+G_K(t)}{2G_\pi(m_V^2)}},\nonumber\\
G_P(t)&=&\sqrt{t}\left(\frac{2 q_P(t)}{\sqrt{t}}\right)^3,
\end{eqnarray}
and $m_V$, $\Gamma_V$ correspond to the masses and widths of
the resonances given in Table~\ref{tab:g11para}.  
Note that $\hat q_P^2(t)\equiv q_P^2(t)\Theta(t-4m_P^2)$ vanishes below 
the $2m_P$ threshold. In particular, Eq.(\ref{ec:g11}) 
below $K\bar K$ threshold is similar  to the
widely used Kuhn and Santamar\'{\i}a form in \cite{Kuhn:1990ad}.
In this region, since the coupling to the 4-pion state is negligible
and $\pi\pi$ scattering is elastic, 
Watson's Theorem implies that $\phi_1^1(t)$ should be equal to 
 the phase shift of the $I=1$, $\ell=1$ partial wave of $\pi\pi$ scattering. 
Since $C$ and $r_1$ are real, they do not contribute to the phase, nor $\beta_1$
nor $\gamma_1$, being multiplied by $\hat q_K^2$, 
so that the parameters $m_\rho, \Gamma_\rho, \beta, \gamma$ are obtained from a fit to 
the dispersive analysis \cite{GarciaMartin:2011cn} of the $\pi \pi$ phase shift in the elastic region. Indeed, in the lower panel of Fig.\ref{fig:g11ufd} 
it can be seen that our parameterization describes remarkably well the 
$\pi\pi$ scattering data on the phase below $K\bar K$ threshold.

The parameters of the $\rho''$ resonance 
are fixed for simplicity to those of the RPP \cite{RPP},
whereas those for the $\rho'$ are allowed to vary within 1.5 standard deviations
within the values listed in the PDG. Note that the ones determined by the CLEO Collaboration
\cite{Anderson:1999ui} are not compatible with our best fit, if one tries to fix those parameters to reproduce the $\pi \pi \rightarrow K \bar K$ data the $\chi^2$ is increased by almost a factor of $2$. 
Then we fit the rest of the parameters to describe the data in the physical and pseudophysical regions, the best result is shown in Fig.~\ref{fig:g11ufd} and the parameters are given in Table \ref{tab:g11para}. The fit has a total $\chi^2/dof=1.7$, but a slightly larger $\chi^2/dof=2.2$ is found in the physical region. Conservatively we use the square root of the latter to rescale 
the fit parameter uncertainties in the table.

\begin{table}[h] 
\caption{Parameters of the $g^1_1$ wave.
Masses and widths are given in GeV whereas, $C$, $\beta_1,\gamma_1$ and $r_1$
are given in GeV$^{-2}$.}
\centering 
\begin{tabular}{c c c } 
\hline\hline  
\rule[-0.05cm]{0cm}{.35cm}Parameter & UFD & CFD \\ 
\hline 
\rule[-0.05cm]{0cm}{.35cm}$m_\rho$ & 0.7757 $\pm$0.0010 & 0.7749 $\pm$0.0010\\
\rule[-0.05cm]{0cm}{.35cm}$\Gamma_\rho$ &0.152 $\pm$0.001  &0.153 $\pm$0.001\\ 
\rule[-0.05cm]{0cm}{.35cm}$m_{\rho'}$ & 1.440$\pm$0.015 & 1.438$\pm$0.015\\
\rule[-0.05cm]{0cm}{.35cm}$\Gamma_{\rho'}$ & 0.310$\pm$0.029  & 0.309$\pm$0.029\\ 
\rule[-0.05cm]{0cm}{.35cm}$m_{\rho''}$ & 1.72 & 1.72 \\
\rule[-0.05cm]{0cm}{.35cm}$\Gamma_{\rho''}$ & 0.25  & 0.25\\ 
\rule[-0.05cm]{0cm}{.35cm}$C$ & 1.21 $\pm$0.11  & 1.23 $\pm$0.11\\ 
\rule[-0.05cm]{0cm}{.35cm}$r_1$ & 3.95 $\pm$0.76  & 3.43 $\pm$0.76\\ 
\rule[-0.05cm]{0cm}{.35cm}$\beta$ & -0.168 $\pm$0.007  & -0.172 $\pm$0.007\\ 
\rule[-0.05cm]{0cm}{.35cm}$\beta_1$ & 0.37 $\pm$0.02  & 0.38 $\pm$0.02\\ 
\rule[-0.05cm]{0cm}{.35cm}$\gamma$ & 0.10 $\pm$0.02  & 0.14 $\pm$0.02\\ 
\rule[-0.05cm]{0cm}{.35cm}$\gamma_1$ & -0.06 $\pm$0.06  & -0.17 $\pm$0.06\\
\hline 
\end{tabular} 
\label{tab:g11para} 
\end{table}

The data and the results of our Unconstrained Fit to Data (UFD) are shown in Fig.~\ref{fig:g11ufd}. Note that we plot the modulus from $K\bar K$ threshold and that, as already commented, data only reaches up to 1.57 GeV. The shape above that energy is almost entirely given by the $\rho''$ resonance. Concerning the phase, from the two-pion threshold to the $K \bar K$ threshold 
it is indistinguishable from that obtained from the $\pi\pi$ dispersive analysis
in \cite{GarciaMartin:2011cn}. In Fig.\ref{fig:g11ufd} our result below threshold can be compared to the data from elastic $\pi\pi$ scattering \cite{Protopopescu:1973sh,Estabrooks:1974vu}.
Note also the large uncertainty of both the data and the error bands in the region around 1.5 GeV, which is due to the fact that  the modulus almost vanishes there. Fortunately, 
this will also make the contribution of that region to the dispersive integrals almost negligible.

\subsubsection{$\ell=2,I=0$ partial wave}
\label{subsec:ufd20}

\begin{figure}
\includegraphics[scale=0.32]{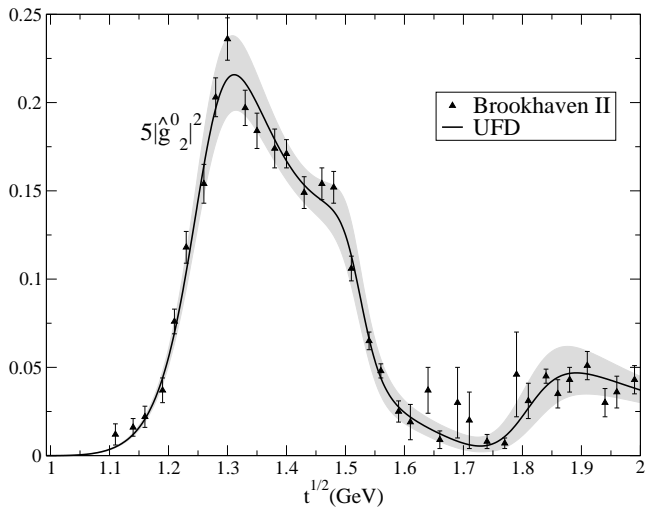}
\caption{\rm \label{fig:g20ufd} Data on the modulus of $\hat g^0_2(t)$
from the Brookhaven-II analysis \cite{Longacre:1986fh} together with our UFD fit,
described in the text.
}
\end{figure}

The data in Fig.~\ref{fig:g20ufd} that we use for this wave in the physical region were obtained in the
Brookhaven-II analysis \cite{Longacre:1986fh},
published 6 years after Brookhaven-I. 
The Brookhaven-II work was a study of 
the $I=0$, $J^{PC}=2^{++}$ channel of
$\pi\pi\rightarrow\bar KK$ scattering
within a coupled channel formalism, which 
included data from other reactions.
The latest Brookhaven-III re-analysis by some members of that collaboration, 
including even further information on other processes 
can be found in \cite{Lindenbaum:1991tq}. 
Note that our normalization differs from that in the experimental works
and this is why we are plotting $\vert \hat g^0_2\vert$, defined as:
\begin{equation}
\hat g^0_2(t)\equiv\frac{2(q_\pi q_K)^{5/2}}{\sqrt{t}} g^0_2(t)\equiv
\vert \hat g^0_2(t) \vert \exp(i\phi^0_2(t)).
\end{equation}

Contrary to the previous $\ell=1,I=1$  case, where the $\rho(770)$ resonance dominates
the unphysical region, now the lowest resonance is well above the $K\bar K$ 
threshold and therefore it does not dominate the unphysical region. 
Thus our $\ell=2,I=0$ parameterization will have two pieces: 
one above $K\bar K$ 
threshold and another one below.

Concerning the physical region, $t\geq t_K$, note that
there are only data for the modulus $\vert \hat g^0_2\vert$, Fig~\ref{fig:g20ufd}.
Therefore, since we also need to have a phase we use 
a  phenomenological
description in terms of resonances
similar to that in \cite{Lindenbaum:1991tq}, 
which is a sum of usual Breit-Wigner 
shapes, although since they overlap significantly we include some interference phases.
We thus use:
\begin{eqnarray}
\hat g^0_2(t)&=&\frac{C\sqrt{(q_\pi(t)q_K(t))^5}}
{\sqrt{t}\sqrt{1+r_2^2 \hat q_\pi^4(t)}\sqrt{1+r_2^2 \hat q_K^4(t)}}\label{eq:g02}\\
&\times& \Big\{ e^{i\phi_1}BW(t)_1 +\beta e^{i\phi_2}BW(t)_2 
+ \gamma e^{i\phi_3}BW(t)_3 \Big\}, \nonumber
\end{eqnarray}
with 
\begin{eqnarray}
BW(t)_T&=&\frac{m^2_T}{m_T^2-t-{\it i}m_T\Gamma_T(t)},\label{eq:BW}\\
\Gamma_T(t)&=&\Gamma_T\left(\frac{q_T(t)}{q_T(m_T^2)}\right)^{5}\frac{m_T}{\sqrt{t}}
\frac{D_{2}(r\, q_T(m_T^2))}{D_{2}(r\,q_T(t))}, \nonumber
\end{eqnarray}
 where $D_{2}(x)=9+3x^2+x^4$
provides  the usual Blatt-Weisskopf barrier factor for $\ell=2$, with 
a typical $r=5\,\gev^{-1}\,\simeq 1\,$fm.

In Eq.\eqref{eq:BW} above, $T=1,2,3$ stands for the tensor $f_2(1270)$, $f'_2(1525)$ and $f_2(1810)$ resonances, respectively. Since they decay predominantly to $\pi\pi$, $\bar KK$
and $\pi\pi$, respectively, we have set $q_1(t)=q_3(t)=q_\pi(t)$, whereas 
$q_2(t)=q_K(t)$. The 
mass $M_T$ and width $\Gamma_T$ of each resonance
after the fit are given in Table~\ref{tab:g20ufd}. As can be seen in the Brookhaven-II and III fits in  \cite{Longacre:1986fh,Lindenbaum:1991tq}, the $f_2'(1525)$ was at odds with the present knowledge about this resonance parameters. Moreover, the parameters of the $f_2(1810)$ vary within a huge range even when using almost the same data. As we have no data for the phase of the partial wave it is not possible to fix the position of the masses with accuracy, however, performing a coupled-channel analysis for the tensor partial wave is out of the scope of this work, mostly because we have no dispersive control over other channels apart from $\pi \pi \rightarrow K \bar{K}$. For that reason we have included the masses of both the $f_2(1270)$ and the $f'_2(1525)$ as additional data for our fit. In particular, we take as input for the fit $m_{f_2}=1.2755 \pm 0.0035\,$GeV which is the average and standard deviation of the values used in the RPP's own average \cite{RPP}. This we do to have a more conservative estimate of the systematic uncertainty. 
For the $f_2'$ we take directly the RPP average $m_{f'_2}=1.525 \pm 0.005\,$GeV.
The inclusion of the $f_2(1810)$  is 
purely phenomenological, following \cite{Longacre:1986fh,Lindenbaum:1991tq}, just to describe the final rise seen in the modulus, 
but this resonance still ``needs confirmation'' according to the RPP. We could have described
this raise equally well with another functional form, although it is also clear that there exist some enhancements of the amplitudes and phases for $\pi \pi \rightarrow \pi \pi$ and $\pi \pi \rightarrow \eta \eta$. Its numerical effect on our dispersive integrals is rather small.
In Table~\ref{tab:g20ufd} we also provide the phases $\phi_T$ resulting from the fit to data.

Concerning the unphysical region, $t<t_K$, since the 
contribution of the four pion state is negligible, we have assumed that 
$\pi\pi$ scattering is elastic. Hence 
we can use Watson's Theorem to identify $\phi_2^0=\delta_2^{(0)}$, 
where  $\delta_2^{(0)}$ is the $\pi\pi$-scattering
phase shift. 
Then we have fitted $\delta_2^{(0)}$ to the result 
obtained in \cite{GarciaMartin:2011cn} from a dispersive analysis of
$\pi\pi$ scattering data. For this we have used a conformal expansion similar to
that in \cite{GarciaMartin:2011cn} but with one more parameter $B_2$ fixed to ensure 
a continuous matching of $g^0_2$ at threshold. Namely:
\begin{eqnarray}
\cot\phi_2^{0}(t)&=&\frac{t^{1/2} }{2q_\pi^5}(m_{f_2(1270)}^2-t)m_\pi^2\times\nonumber\\
&&\quad\left\{
B_0+B_1 w(t)+B_2w(t)^2\right\},
\nonumber\\
w(t)&=&\frac{\sqrt{t}-\sqrt{t_0-t}}{\sqrt{t}+\sqrt{t_0-t}},\quad
t_0^{1/2}=1.05 \, \gev, \quad
  \label{eq:D0lowparam} 
\end{eqnarray}
where
\begin{equation}
  B_2\,\omega(t_K)^2=\frac{q_\pi^5(t_K)\cot(\phi^0_2(t_K))}{m_K(m_{f_2(1270)}^2-t_K)m_\pi^2} -B_0-B_1\,\omega(t_K) ,
\end{equation}
has been fixed by continuity with the piece above $t_K$ in Eq.\eqref{eq:g02}.
In Table~\ref{tab:g20ufd}  we provide values of 
$B_0, B_1$ after fitting the CFD phase-shift in \cite{GarciaMartin:2011cn}.
With this parameterization we obtain a final $\chi^2/dof=1.4$. Thus we rescale our uncertainties by a factor of $\sim 1.2$.
We have checked that this phase is also compatible within uncertainties
with the dispersive analysis of the $\pi \pi$ D-wave 
using Roy and GKPY equations in \cite{Bydzovsky:2016vdx}.

\begin{table}[h] 
\caption{Parameters of the $g^0_2$ wave.} 
\centering 
\begin{tabular}{c c c } 
\hline\hline  
\rule[-0.05cm]{0cm}{.35cm}Parameter & UFD & CFD \\ 
\hline 
\rule[-0.05cm]{0cm}{.35cm}$m_{f_2(1270)}$ & 1.271 $\pm$0.0035GeV  & 1.271 $\pm$0.0035GeV\\
\rule[-0.05cm]{0cm}{.35cm}$m_{f'_2(1525)}$ & 1.522 $\pm$0.005 GeV  & 1.522 $\pm$0.005 GeV\\ 
\rule[-0.05cm]{0cm}{.35cm}$m_{f_2(1810)}$ & 1.806 $\pm$0.017 GeV  & 1.802 $\pm$0.017 GeV\\ 
\rule[-0.05cm]{0cm}{.35cm}$\Gamma_{f_2(1270)}$ & 0.187 $\pm$0.009 GeV  & 0.191 $\pm$0.009 GeV\\
\rule[-0.05cm]{0cm}{.35cm}$\Gamma_{f'_2(1525)}$ & 0.108 $\pm$0.016 GeV  & 0.107 $\pm$0.016 GeV\\
\rule[-0.05cm]{0cm}{.35cm}$\Gamma_{f_2(1810)}$ & 0.201 $\pm$0.028 GeV  & 0.198 $\pm$0.028 GeV\\
\rule[-0.05cm]{0cm}{.35cm}$\phi_{f_2(1270)}$ & -0.049 $\pm$0.014 & -0.078 $\pm$0.014 \\
\rule[-0.05cm]{0cm}{.35cm}$\phi_{f'_2(1525)}$ & 2.62 $\pm$0.16  & 2.59 $\pm$0.16\\
\rule[-0.05cm]{0cm}{.35cm}$\phi_{f_2(1810)}$ & -0.72$\pm$0.16  & -0.82$\pm$0.16  \\
\rule[-0.05cm]{0cm}{.35cm}$B_0$ & 12.5 $\pm$ 0.4 & 12.4 $\pm$ 0.4  \\
\rule[-0.05cm]{0cm}{.35cm}$B_1$ & 10.3 $\pm$ 1.0 & 12.3 $\pm$ 1.0  \\
\rule[-0.05cm]{0cm}{.35cm}$C$ & 1.82 $\pm$ 0.09 GeV$^{-2}$ & 1.86 $\pm$ 0.09  GeV$^{-2}$\\
\rule[-0.05cm]{0cm}{.35cm}$r_2^2$ & 6.68 $\pm$ 0.72 GeV$^{-4}$ & 6.78 $\pm$ 0.72 GeV$^{-4}$ \\
\rule[-0.05cm]{0cm}{.35cm}$\beta$ & 0.070 $\pm$ 0.016 & 0.066 $\pm$ 0.016  \\
\rule[-0.05cm]{0cm}{.35cm}$\gamma$ & 0.093 $\pm$ 0.02 & 0.094 $\pm$ 0.02  \\
\hline 
\end{tabular} 
\label{tab:g20ufd} 
\end{table}

\begin{figure}
\includegraphics[scale=0.32]{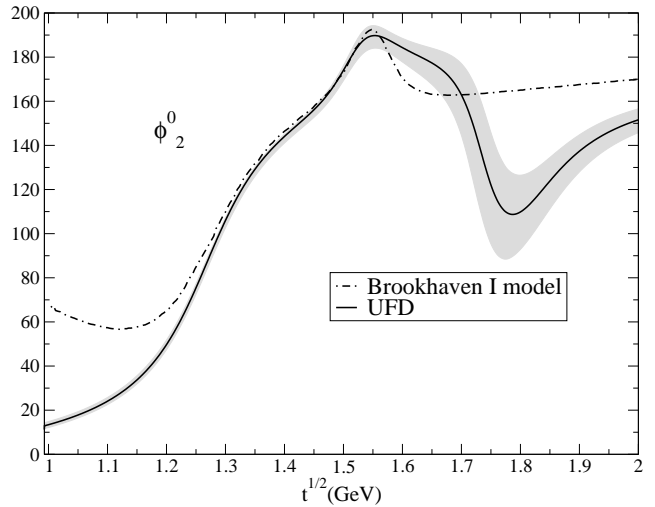}
\caption{\rm \label{fig:comparisong02} Comparison between the UFD $g^0_2$ phase and the one obtained with the Brookhaven-I model.
Note that the latter violates Watson's Theorem at $K\bar K$ threshold.
Also, the former includes an $f_0(1810)$ resonance whereas the latter uses a flat background. 
As explained in the text, the latter is strongly disfavored when fitting Brookhaven II data on the modulus.
}
\end{figure}

Neither Brookhaven-I nor Argonne provide data for this wave, 
nor the models they used to parameterize it. 
Nevertheless Brookhaven-I shows a plot with the central value 
of their phase for this channel, which is later used to extract the $g^0_0$ phase.
As seen in Fig.\ref{fig:comparisong02} our phase is fairly compatible with
the Brookhaven-I model between 1.25 and  1.54 GeV. However,
also in that figure it 
can be seen that the Brookhaven-I model
violates Watson's Theorem at low energies, which our phase fully satisfies. 
In addition,
above 1.6 GeV our phase, obtained by fitting the Brookhaven-II data \cite{Longacre:1986fh}
on the modulus with modern values for the $f_2$ family of resonances, 
is rather different from the flat behavior of the
Brookhaven-I model \cite{Etkin:1981sg} up to 1.9 GeV. 
The reason is that the  Brookhaven-I model used a simple smooth 
background to describe the 1.6-1.9 GeV region, 
instead of the $f_2(1810)$ used in this work.
Actually, we have checked that 
if we impose the phase of the Brookhaven-I model on our fit to the Brookhaven-II
modulus, the resulting $\chi^2 /dof$ is $\sim5$, 
and thus strongly disfavored with respect to our phase.
Even by deforming our fits by including more parameters, 
the best we have been able to achieve
when imposing the phase of the Brookhaven-I model above 1.6 GeV,
is $\chi^2/dof \sim 3$, but at the price of introducing
 contributions difficult to interpret in terms of resonance parameters.
Both the violation of Watson's Theorem and the use of such non-resonant background
make the Brookhaven-I solution suspicious.

Unfortunately the Brookhaven-I model was used to extract the phase of the $g^0_0$,
which therefore also becomes suspicious below 1.2 GeV and above 1.6 GeV.
Nevertheless, and with this caveats in mind
we will still study the $g^0_0$ phase coming 
from the Brookhaven-I collaboration above 1.6 GeV.
The reason is that 
this region lies outside the applicability range of Roy-Steiner equations,
 so that for our purposes is just input. Fortunately, the modulus 
there is very small, so that the contribution from this region 
to the Roy-Steiner equations below 1.6 GeV is very suppressed. 
In Appendix \ref{sec:modifiedg00}, we have checked that either with our $g^0_0$ phase or the Brookhaven-I phase,
the difference lies within our uncertainties in the region up to 1.47 GeV, which is the one of interest for this work 
since it is the one where partial-wave dispersion relations can be applied.

\subsubsection{$\ell=0,I=0$ partial wave}

\begin{figure*}
\centering
\includegraphics[width=.9\textwidth]{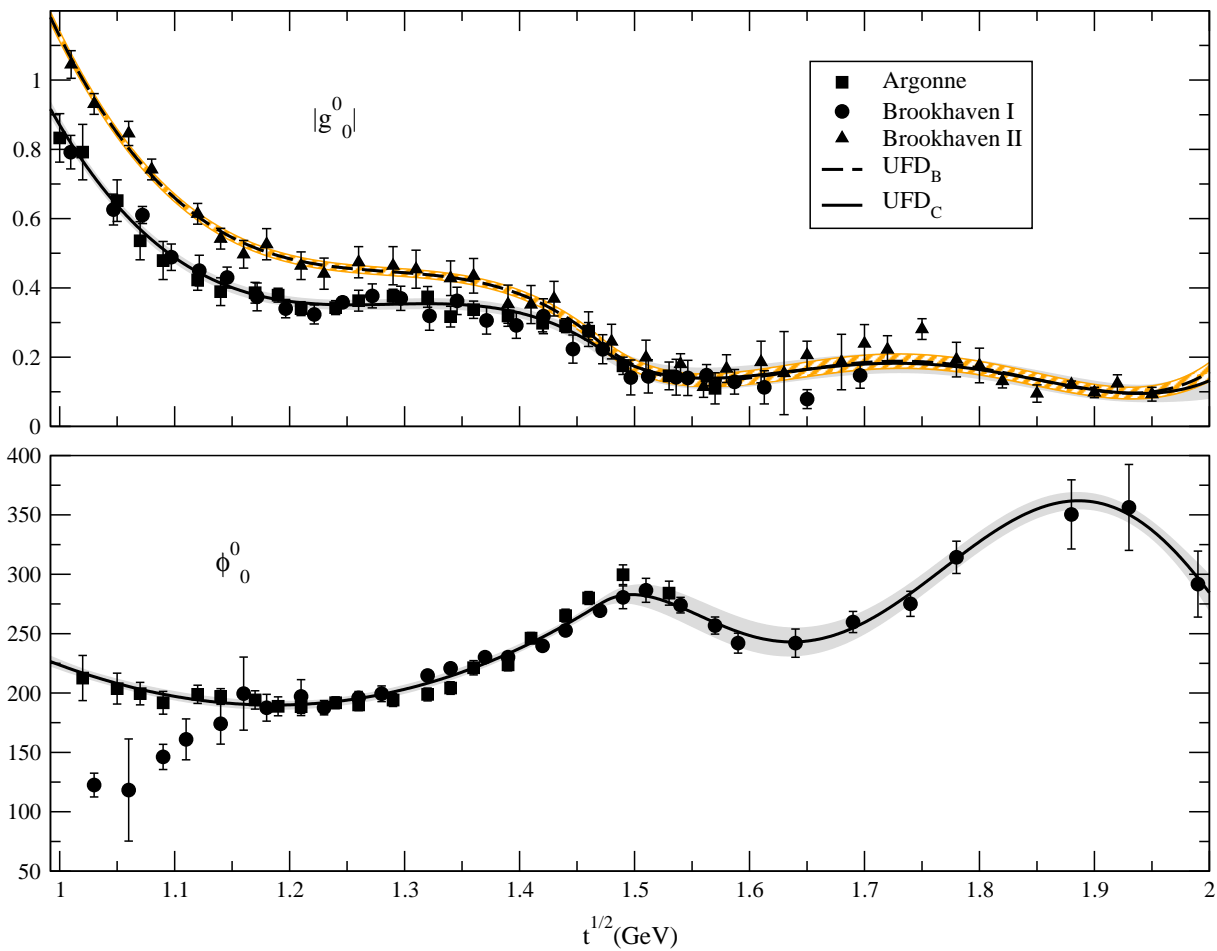}
\caption{ Upper panel: Modulus of the scalar-isoscalar 
$\pi\pi\rightarrow K \bar K$ scattering. The continuous line represents the UFD$_\text{C}$ parameterization while 
the dashed line represents the UFD$_\text{B}$ fit to the Brookhaven-II data only.
Lower panel: Scalar-isoscalar UFD phase for $\pi\pi\rightarrow K \bar K$ scattering, which is common for both UFD$_\text{B}$ and UFD$_\text{C}$.
Note that the Brookhaven-I phase close to threshold lies around 150$^\degree$ or below,
at odds with all dispersive analysis of $\pi\pi$ scattering, which find a phase 
around or above 200$^\degree$.}
 \label{fig:g00ufd} 
\end{figure*}

This wave is the most complicated but also the most interesting 
one for hadron spectroscopy, since here we can find the much debated scalar-isoscalar 
resonances.
For the $g^0_0(t)$ partial wave there are data in the whole region of interest 
on both the modulus $\vert g_0^0\vert$ and the phase
$\phi^0_0$, 
which we show in  Fig.\ref{fig:g00ufd}. 
The data sets extend up to 2.4 GeV, but we do not fit that region
because from 2 GeV we will use Regge parameterizations.
It is then convenient to  split into two regions the data description below 2 GeV:

\begin{enumerate}
\item[I)] Region I: From $\sqrt{t_{min,I}}=2m_K$ up to $\sqrt{t_{max,I}}=1.47\,$GeV, where data from Argonne \cite{Cohen:1980cq}
and Brookhaven-I \cite{Etkin:1981sg} coexist. Note that  
this region will lie within the applicability of Roy-Steiner equations and 
will be later constrained to satisfy dispersion relations.

Concerning the phase  $\phi^0_0$, it is clearly seen in Fig.\ref{fig:g00ufd}
that from $2m_K$ up to $1.2\,$GeV, the Argonne \cite{Cohen:1980cq} and Brookhaven-I
\cite{Etkin:1981sg} sets are incompatible. 
Let us now recall that, by Watson's Theorem, $\phi^0_0$
at $K \bar{K}$ threshold should match the scalar-isoscalar
$\pi\pi\rightarrow\pi\pi$ phase shift $\delta_0^{(0)}$. 
However, the $\pi\pi$ scattering analyses 
with Roy and GKPY equations 
that extend up to or beyond $K \bar K$ threshold \cite{GarciaMartin:2011cn,Moussallam:2011zg} 
find $\delta_0^{(0)}>200^\degree$, which is consistent with the
Argonne \cite{Cohen:1980cq} phase, but much higher than  
the phase of Brookhaven-I \cite{Etkin:1981sg}. 
In addition we have just seen that this phase was extracted using a $g_2^0$ wave that also violates Watson's Theorem.
Therefore, for our fits we have discarded the phase of Brookhaven-I 
\cite{Etkin:1981sg} below $\sim$1.15 GeV, i.e. until it agrees with that of 
Argonne \cite{Cohen:1980cq}. 

Concerning the data on $\vert g_0^0\vert$, 
shown in Fig.~\ref{fig:g00ufd}, the 
Argonne and Brookhaven-I sets are consistent among themselves
but not with the Brookhaven-II. 
However, the latter is consistent up to 1.2 GeV with the dip solution 
for the inelasticity favored from dispersive
analyses of $\pi\pi\rightarrow\pi\pi$ scattering \cite{GarciaMartin:2011cn,Moussallam:2011zg} 
(assuming that only $\pi\pi$ and $K\bar K$ states are relevant). 
Finally, the ``dip'' solution from $\pi\pi$ scattering in the 1.2 GeV to 1.47 region has such large uncertainties that
is roughly consistent with the three data sets.

\item[II)] 
In the region from $\sqrt{t_{min,II}}=1.47$ GeV to $\sqrt{t_{max,II}}=2$ GeV Roy-Steiner equations will not be applicable and
thus this region 
will only be used as input for our dispersive calculations for lower energies.
Note that here all experiments are roughly consistent,
although the Argonne set only reaches up to $\sim$1.5 GeV, Brookhaven-I up to $\sim$1.7 GeV
and only Brookhaven-II reaches up to 2 GeV.
\end{enumerate}

Therefore in order to test different data sets independently
and to be able to impose later Roy-Steiner equations as constraints below 1.5 GeV 
using as input the region above, we have decided to parameterize our 
amplitudes by piecewise functions.
Actually, each piece will be parameterized by
 Chebyshev polynomials, because they are rather simple and, in practice,
tend to reduce the correlation between the small number of parameters needed 
to obtain a good fit. They are given by:
\begin{eqnarray}
&&p_0(x)=1,\quad p_1(x)=x,     \nonumber\\
&&p_{n+1}(x)=2xp_n(x)-p_{n-1}(x).
\end{eqnarray}
Thus we first map each energy region  $i=I,II$ 
into the $x\in[-1,1]$ interval through the lineal transformation 
\begin{equation}
x_i(t)=2\frac{\sqrt{t}-\sqrt{t_{min,i}}}{\sqrt{t_{max,i}}-\sqrt{t_{min,i}}}-1.
\end{equation}
Note that for any $n$, $p_n(1)=1$ and $p_n(-1)=(-1)^n$,
which is useful for  matching the different pieces
smoothly up to the first derivative.

Since for the $\phi_0^0$ phase we have already selected a single set on each region, 
our Unconstrained Fit to Data (UFD) will be given in just two pieces: 
\begin{equation} \phi_0^0(t)=\left\{
\begin{array}{@{}rl@{}}
\sum^3_{n=0}{B_n p_n(x_I(t))}, & \text{Region I,}\\
 & \\
\sum^5_{n=0}{C_n p_n(x_{II}(t))},&\text{Region II.}
\end{array}
\right.
\end{equation}  
Note that we set:
\begin{align}
&B_0=\delta_0^{(0)}(t_K)+B_1-B_2+B_3,\\
&C_0=\phi_0^0(t_{max,I})+C_1-C_2+C_3-C_4+C_5,
\end{align}
in order to impose continuity at $K\bar K$ threshold and between the two energy regions, respectively.
In addition, we fix $C_1$ to have a continuous derivative 
for the central value of the curve and we take $\delta_0^{(0)}(t_K)=(226.5\pm1.3)^\degree$ from \cite{GarciaMartin:2011cn}. The rest of the parameters of the fit are given in Table~\ref{tab:g00phase}. The total $\chi^2/dof=1.47$, which comes slightly larger than one due to some incompatibilities between data sets. 
Consequently, the
uncertainties of the parameters in Table~\ref{tab:g00phase} have been rescaled by a factor $\sqrt{1.5}$.

\begin{table}[h] 
\caption{Parameters of $\phi^0_0$.} 
\centering 
\begin{tabular}{c c c c} 
\hline\hline  
\rule[-0.05cm]{0cm}{.35cm}Parameter & UFD & CFD$_\text{B}$ & CFD$_\text{C}$ \\ 
\hline 
\rule[-0.05cm]{0cm}{.35cm}$B_1$ & 23.6  $\pm$1.3  & 22.1 $\pm$1.3 & 22.9  $\pm$1.3\\ 
\rule[-0.05cm]{0cm}{.35cm}$B_2$ & 29.4  $\pm$1.3  & 27.7  $\pm$1.3 & 28.4  $\pm$1.3\\ 
\rule[-0.05cm]{0cm}{.35cm}$B_3$ & 0.6  $\pm$1.6  & 1.8 $\pm$1.6 &   1.1  $\pm$1.6\\ 
\rule[-0.05cm]{0cm}{.35cm}$C_1$ & 34.3932 fixed  & 35.3450 fixed & 34.51593 fixed\\ 
\rule[-0.05cm]{0cm}{.35cm}$C_2$ & 4.4  $\pm$2.6  & 4.3  $\pm$2.6 & 4.3  $\pm$2.6\\ 
\rule[-0.05cm]{0cm}{.35cm}$C_3$ &-32.9 $\pm$5.2  &-33.3 $\pm$5.2 &-32.6 $\pm$5.2\\ 
\rule[-0.05cm]{0cm}{.35cm}$C_4$ &-16.0 $\pm$2.2  &-16.5 $\pm$2.2 &-16.0 $\pm$2.2\\ 
\rule[-0.05cm]{0cm}{.35cm}$C_5$ &  7.4 $\pm$2.4  &  7.2 $\pm$2.4 &  7.2 $\pm$2.4\\ 
\hline 
\end{tabular} 
\label{tab:g00phase} 
\end{table}

In contrast, for the modulus
 we want to test different sets of data.
Thus, we have performed two Unconstrained Fits to Data (UFD) in Region I:
i)  A UFD$_\text{B}$ fitting the data of Brookhaven-II \cite{Longacre:1986fh}.
ii) A UFD$_\text{C}$ fitting the 
``Combined''   data of Argonne  \cite{Cohen:1980cq} and Brookhaven-I \cite{Etkin:1981sg}.
Both use the same data in Region II. 
Thus we will use the following functional form: 
\begin{equation} 
\vert g_0^0(t)\vert=\left\{
\begin{array}{@{}rl@{}}
\sum^3_{n=0}{D_n p_n(x_I(t))}, & \text{Region I,}\\
 & \\
\sum^4_{n=0}{F_n p_n(x_{II}(t))},&\text{Region II,}
\end{array}
\right.
\label{eq:ufd}
\end{equation}  
where we now set:
\begin{equation}
F_0=\vert g_0^0(t_{max,I})\vert+F_1-F_2+F_3-F_4, 
\end{equation}
in order to ensure continuity between the two regions
and we fix $F_1$ to ensure a continuous derivative for the central value.


Both the UFD$_\text{B}$ and UFD$_\text{C}$ fits, whose parameters are given in Tables~\ref{tab:ufdb} and \ref{tab:ufdc}, respectively, 
 have $\chi^2/dof\sim 1$
and are shown in the upper panel of Fig.~\ref{fig:g00ufd}.

\begin{table}[h] 
\caption{Parameters of the UFD$_\text{B}$ and CFD$_\text{B}$ fits to $\vert g^0_0\vert $. } 
\centering 
\begin{tabular}{c c c } 
\hline\hline  
\rule[-0.05cm]{0cm}{.35cm}Parameter & UFD$_\text{B}$ & CFD$_\text{B}$ \\ 
\hline 
\rule[-0.05cm]{0cm}{.35cm}$D_0$ & 0.59 $\pm$0.01  & 0.60 $\pm$0.01\\
\rule[-0.05cm]{0cm}{.35cm}$D_1$ &-0.38 $\pm$0.01  &-0.35 $\pm$0.01\\ 
\rule[-0.05cm]{0cm}{.35cm}$D_2$ & 0.12 $\pm$0.01  & 0.13 $\pm$0.01\\ 
\rule[-0.05cm]{0cm}{.35cm}$D_3$ &-0.09  $\pm$0.01  &-0.12  $\pm$0.01\\ 
\rule[-0.05cm]{0cm}{.35cm}$F_1$ &-0.04329  fixed  &-0.04078  fixed\\
\rule[-0.05cm]{0cm}{.35cm}$F_2$ &-0.008  $\pm$0.009  &-0.007  $\pm$0.009\\ 
\rule[-0.05cm]{0cm}{.35cm}$F_3$ &-0.028  $\pm$0.007  &-0.035  $\pm$0.007\\ 
\rule[-0.05cm]{0cm}{.35cm}$F_4$ & 0.026  $\pm$0.007  & 0.037  $\pm$0.007\\ 
\hline 
\end{tabular} 
\label{tab:ufdb} 
\end{table}

\begin{table}[h] 
\caption{Parameters of the UFD$_\text{C}$ and CFD$_\text{C}$ fits to $\vert g^0_0\vert $. }
\centering 
\begin{tabular}{c c c } 
\hline\hline  
\rule[-0.05cm]{0cm}{.35cm}Parameter & UFD$_\text{C}$ & CFD$_\text{C}$ \\ 
\hline 
\rule[-0.05cm]{0cm}{.35cm}$D_0$ & 0.46 $\pm$0.01  & 0.46 $\pm$0.01\\
\rule[-0.05cm]{0cm}{.35cm}$D_1$ &-0.27 $\pm$0.01  &-0.25 $\pm$0.01\\ 
\rule[-0.05cm]{0cm}{.35cm}$D_2$ & 0.11 $\pm$0.01  & 0.11 $\pm$0.01\\ 
\rule[-0.05cm]{0cm}{.35cm}$D_3$ &-0.078  $\pm$0.009  &-0.087  $\pm$0.009\\ 
\rule[-0.05cm]{0cm}{.35cm}$F_1$ &-0.04153 fixed   &-0.03738 fixed \\
\rule[-0.05cm]{0cm}{.35cm}$F_2$ &-0.010  $\pm$0.008  &-0.013  $\pm$0.008\\ 
\rule[-0.05cm]{0cm}{.35cm}$F_3$ &-0.023  $\pm$0.007  &-0.025  $\pm$0.007\\ 
\rule[-0.05cm]{0cm}{.35cm}$F_4$ & 0.021  $\pm$0.006  & 0.025  $\pm$0.006\\ 
\hline 
\end{tabular} 
\label{tab:ufdc} 
\end{table}

\subsubsection{Partial waves with $\ell>2$}
\label{sec:higherwaves}

For higher partial waves we just use Breit-Wigner descriptions associated to the poles listed in the PDG.
In particular, for the $g^1_3(t)$ we include a single $\rho_3(1690)$ resonance. 
The $\ell=4$ partial wave, parameterized as an $f_2(2050)$ Breit-Wigner resonance, is only included in the $g^0_2(t)$ dispersive calculation due to 
its negligible contribution below 2 GeV for the $g^0_0(t)$.

\subsection{Higher energies}
\label{sec:Regge}

There is no high-energy experimental information on $\pi\pi\rightarrow \bar KK$ nor
$\pi K\rightarrow \pi K$. 
However, the high energy behavior of both processes can be confidently 
modeled by applying factorization
to Regge amplitudes obtained for other processes.
In this work we will use, for the $s$-channel above 1.74 GeV the 
Regge model description presented in \cite{Pelaez:2003ky} and updated in \cite{GarciaMartin:2011cn,Pelaez:2016tgi},
whereas for the $t$-channel we will use the asymptotic forms of 
the Veneziano model \cite{Veneziano:1968yb}, with the updated parameters in 
\cite{Buettiker:2003pp}, to describe 
the process above $2$ GeV. 
The reasons to choose 2 GeV in this work are twofold: on the one hand data for the $g_0^0$ and $g_2^0$ waves reach 
above that energy, on the other hand, even if the $g_1^1$ data end at 1.6 GeV, the $\rho''(1720)$ is well established in the RPP
and with its 250 MeV width, reaches well above 2 GeV. Thus we rely on our partial-wave parameterizations up to 2 GeV, 
but not much more.

In what follows we provide the detail of these descriptions 
using the notation of this work.

For the symmetric amplitude we have the Pomeron $P(s,t)$ contribution 
and the $f_2$ or $P'(s,t)$ exchange:
\begin{eqnarray}
\im F^{+}_{\pi K}(s,t)&=&
\frac{\im F^{(I_t=0)}_{\pi K}(s,t)}{\sqrt{6}}\nonumber\\
&=&\frac{4\pi^2}{\sqrt{6}} f_{ K/\pi}\left[P(s,t)+rP'(s,t)\right],
\label{eq:reggef+}
\end{eqnarray}
where, as explained in \cite{Pelaez:2003ky},
$f_{K/\pi}$ is the factorization that allows to convert one 
$\pi\pi-$Reggeon into a $KK-$Reggeon vertex, whereas $r$ is related to the branching
ratio of the $f_2(1270)$ resonance to $\bar KK$. 
In addition
\begin{eqnarray}
P(s,t)&=&\beta_P\psi_P(t)\alpha_P(t)\frac{1+\alpha_P(t)}{2}e^{\hat b t}\left(\frac{s}{s'}\right)^{\alpha_P(t)},\nonumber\\
P'(s,t)&=&\beta_{P'}\psi_{P'}(t)\frac{\alpha_{P'}(t)(1+\alpha_P(t))}{\alpha_{P'}(0)(1+\alpha_P(0))}e^{\hat bt}\left(\frac{s}{s'}\right)^{\alpha_{P'}(t)},\nonumber\\
\alpha_P(t)&=&1+t\alpha'_P, \psi_P=1+c_Pt,\nonumber\\
\alpha_{P'}(t)&=&\alpha_{P'}(0)+t\alpha'_{P'}, \psi_P=1+c_{P'}t.
\label{eq:Pomeron}
\end{eqnarray}

In contrast, the antisymmetric amplitude is dominated by just one
contribution coming from the exchange of a Reggeized $\rho$:
\begin{eqnarray}
\im F^{-}_{\pi K}(s,t)&=&\frac{\im F^{(I_t=1)}_{\pi K}(s,t)}{2}\nonumber\\
&=&2\pi^2 g_{ K/\pi}\im T^{(I_t=1)}_{\pi\pi}(s,t),
\label{eq:reggef-}
\end{eqnarray}
where now $g_{ K/\pi}$ is the factorization constant to change a $\pi\pi\rightarrow \rho$ Regge
vertex into $K\bar K\rightarrow \rho$, and 
\begin{eqnarray}
\im T^{(I_t=1)}_{\pi \pi}(s,t)&=& \beta_{\rho}\frac{1+\alpha_{\rho}(t)}{1+\alpha_{\rho}(0)}\varphi(t)e^{\hat bt}\left(\frac{s}{s'}\right)^{\alpha_{\rho}(t)},\nonumber\\
\alpha_{\rho}(t)& =&\alpha_{\rho}(0)+t\alpha'_{\rho}+\frac{1}{2}t^2\alpha''_{\rho},\nonumber\\
\varphi(t)& =&1+d_{\rho}t+e_{\rho}t^2.
\label{eq:reggerho}
\end{eqnarray}

All the parameters in Eqs.~\eqref{eq:Pomeron} and \eqref{eq:reggerho} 
correspond to Regge exchanges without strangeness (the Pomeron, $f_2$ and $\rho$) and 
can be determined \cite{Pelaez:2003ky} from processes that do not involve kaons.
Therefore in this work we fix them, 
both for the unconstrained (UFD) and constrained fits (CFD) here, 
to their updated values of the CFD fits given in \cite{GarciaMartin:2011cn},
 which are listed in Table~\ref{tab:regge}.
Let us remark that with these parameters our asymptotic
value of the Pomeron $\pi K$ cross section 
is $\simeq 10.3\,$mb.
This is about twice the $\simeq 5\pm2.5\,$mb value used in 
\cite{Buettiker:2003pp}. This value was inspired
by the work in \cite{Ananthanarayan:2000ht}, 
which asymptotically yielded $6\pm 5$ mb for  $\pi\pi$ 
scattering. However, this $\pi\pi$ value has been 
revisited recently by members of the same group \cite{Caprini:2011ky} 
yielding $12.2\pm 0.1\,$mb for $\pi\pi$ scattering, thus supporting our larger value for $\pi K$ rather than $5\pm2.5\,$mb.

\begin{table}[h]
\caption{Values of Regge parameters obtained in \cite{Pelaez:2004vs,GarciaMartin:2011cn}. Since these could be fixed using reactions other than $\pi K$ scattering,
they will be fixed both in our UFD and CFD parameterizations.
\label{tab:regge} } 
\centering 
\begin{tabular}{c c} 
\hline\hline  
\rule[-0.15cm]{0cm}{.55cm} Regge & Used both for \\ 
\rule[-0.15cm]{0cm}{.55cm} Parameters & UFD and CFD \\ 
\hline 
$s'$ & 1 GeV$^{2}$                                  \\
$\hat b$ & 2.4                 $\pm$0.5 GeV$^{-2}$       \\ 
$\alpha'_{P}$ & 0.2       $\pm$0.1 GeV$^{-2}$       \\
$\alpha'_{P'}$ & 0.9 GeV$^{-2}$                     \\ 
$c_{P}$ & 0.6             $\pm$1 GeV$^{-2}$         \\
$c_{P'}$ & -0.38          $\pm$0.4 GeV$^{-2}$       \\ 
$\beta_{P}$ & 2.50        $\pm$0.04                 \\
$c_{P}(0)$ & 0            $\pm$0.04                 \\
$\beta_{P'}$ & 0.80       $\pm$0.05                 \\
$c_{P'}(0)$ & -0.4        $\pm$0.4                  \\
$\alpha_{P'}(0)$ & 0.53   $\pm$0.02                 \\
$\alpha_{\rho}(0)$ & 0.53 $\pm$0.02                 \\ 
$\alpha'_{\rho}$ & 0.9 GeV$^{-2}$                   \\
$\alpha''_{\rho}$ & -0.3 GeV$^{-4}$                 \\ 
$d_{\rho}$ & 2.4          $\pm$0.5 GeV$^{-2}$       \\
$e_{\rho}$ & 2.7          $\pm$2.5                  \\
$\beta_{\rho}$ & 1.47     $\pm$0.14                 \\
\hline 
\end{tabular} 
\end{table} 
In contrast, the determination of the
parameters $f_{K/\pi}$, $r$ and $g_{K/\pi}$ needs input from kaon interactions.
In principle all them were 
determined in \cite{Pelaez:2003ky} from $KN$ factorization
and we take the $f_{K/\pi}$ and $r$ values from that reference. Concerning
$g_{K/\pi}$ we take the updated value from the Forward Dispersion 
Relation study of $\pi K$ scattering in 
\cite{Pelaez:2016tgi} (we use the 
value from the CFD there).
Their values can be found in Table~\ref{tab:reggepiK}. Since their determination 
involves kaon interactions, we will allow them to vary when constraining
our fits with dispersion relations, i.e. from the UFD to the CFD sets.
However, in the table it is seen that the change is minute.

\begin{table}[h]
\caption{Values of Regge parameters 
involving strangeness. They are all allowed to vary from our UFD to our CFD sets with the exception of $\alpha_{K^*}$ and $\alpha'_{K^*}$, since they are both determined from linear Regge trajectory fits to strange resonances.
\label{tab:reggepiK} } 
\centering 
\begin{tabular}{c c c} 
\hline\hline  
\rule[-0.15cm]{0cm}{.55cm} Regge & UFD & CFD\\ 
\hline           
$f_{K/\pi}$  &  0.66 fixed     &  0.66 fixed      \\
$g_{K/\pi}$  &  0.53 fixed     &  0.53 fixed      \\
$r$          &  0.05$\pm$0.010     &  0.052$\pm$0.010     \\ 
\hline           
$\alpha_{K^*}$     &  0.352  &  0.352                  \\
$\alpha'_{K^*}$    &  0.882 GeV$^{-2}$  &  0.882 GeV$^{-2}$      \\ 
$\lambda$    &  11.0$\pm$5.0          &  10.7$\pm$5.0      \\
\hline 
\end{tabular} 
\end{table} 

For the $t$-channel, $\pi\pi\rightarrow K\bar K$,
 we also need the exchange of strange Reggeons, for which we will 
assume that the dominant trajectories $K^*_1(892)$ and $K_2^*(1430)$ are degenerate,
Thus we use for them a common trajectory 
$\alpha_{K^ *}(s)=\alpha_{K^ *}+\alpha'_{K^ *}s$
whose parameters, listed in Table~\ref{tab:reggepiK}, are obtained from the linear Regge 
trajectories for strange resonances and therefore are kept fixed
for both our UFD and CFD sets.

All these features are nicely incorporated in the
 dual-resonance Veneziano-Lovelace 
model \cite{Veneziano:1968yb,pipibook},
which was already used in the Roy-Steiner context for $\pi K$ scattering
\cite{Ananthanarayan:2001uy}. 
Here we are only interested in the asymptotic behavior
\cite{Buettiker:2003pp}:
\begin{align}
&\frac{\im G^0(t,s_b)}{\sqrt{6}}\Big|_{\rm Regge}=\frac{\im G^1(t,s_b)}{2}\Big|_{\rm Regge}= \nonumber \\
&\frac{\pi\lambda(\alpha'_{K^*}t)^{\alpha_{K^*}+a \alpha'_{K^*}}}{\Gamma(\alpha_{K^*}+a \alpha'_{K^*})}
\Big[1+\frac{\alpha'_{K^*} b}{t}(\psi(\alpha_{K^*}+a \alpha'_{K^*}) \nonumber \\
&-\log(\alpha'_{K^*} t)\Big)\Big],
\label{eq:vene}
\end{align}
where $\psi$ is the polygamma function. Note that the $a, b$ parameters in the above equation
will be those defining the hyperbola $(s-a)(u-a)=b$ along which we will define our hyperbolic dispersion relations in the next section. For a given $t$, $s_b$ is the value of $s$ that lies in the previous hyperbola. In order to compare
with the expressions in \cite{Buettiker:2003pp}, where $a=0$, we have kept just 
the first order in the $b/t$ expansion, although its numerical effect is rather small.

We estimate the remaining $\lambda$ parameter  from 
exact degeneracy between the $\rho$ and $K^*$ families.
We thus match  Eq.\eqref{eq:reggef-} at 2 GeV with
the expression from the degenerate Veneziano model with
its original parameter $\alpha_\rho^V=0.475$.
In this way we find 
\begin{equation}
\lambda\simeq 
\frac{2\pi\Gamma(\alpha_\rho^V)}{\alpha'^{\alpha_{\rho}^V}_{K^*}}
 4^{\alpha_\rho-\alpha_\rho^V}\simeq 10.6\pm2.5,
\label{eq:lambdaestimate}
\end{equation}
which is compatible with the value used in \cite{Buettiker:2003pp}, $\lambda=14\pm 5$. 
Conservatively we also add a 25\% uncertainty due to the breaking of degeneracy and thus we arrive to our final estimate 
\begin{equation}
\lambda \simeq 11\pm5,
\label{eq:lambda}
\end{equation}
which for completeness is also listed in Table~\ref{tab:reggepiK}.
Given that it is a crude estimate we will allow this value to vary when constraining our fits to obtain the CFD sets.
We will see that after 
imposing the dispersive constraints we obtain
$\lambda=10.7$, which due to the degeneracy between the $\rho$ and $K^*$ families, 
suggests $g_{K/ \pi}\sim0.55$, 
in perfect agreement with the
value used here that comes from a dispersive $\pi K$ study.

A final remark on 
the size of Regge contributions is in order.
As commented in the introduction, 
in the next sections we will obtain partial-wave dispersion relations
by integrating hyperbolic dispersion relations. 
This is an integral over $b$ for a family of $(s-a)(u-a)=b$ hyperbolas, 
while $a=-10.8 M_\pi^2$ is fixed to the value that maximizes the applicability region (see Appendix~\ref{app:convergence}). 
This means that the exponent $\alpha_{K^*}+a \alpha'_{K^*}<\alpha_{K^*}$
and thus the Regge contribution to $\pi\pi\rightarrow\bar KK$ in this
work, for the same number of subtractions,
is suppressed with respect to its size in \cite{Buettiker:2003pp}, where $a=0$.
This will allow us to consider 
less subtractions without Regge contributions growing large.

\section{Hyperbolic Dispersion Relations and sum rules}
\label{sec:HDR}

Our goal is to calculate a set of parameterizations that describe the 
data up to 1.47 GeV consistently with hyperbolic dispersion relations (HDR). 
As already advanced in the introduction, in this work we will consider a set 
of hyperbolas $(s-a)(u-a)=b$ and use $a$ to maximize the energy domain where the hyperbolic 
dispersion relations hold. 
Note that the phenomenology of the $\pi \pi \rightarrow K \bar{K}$
$a=0$ case has been studied in detail 
in \cite{Ananthanarayan:2001uy,Buettiker:2003pp}. 
Moreover, HDR with $a=0$ were also used for
the study of the $K_0^*(800)$ resonance \cite{DescotesGenon:2006uk}.

In addition, we will use the smallest number of subtractions needed for each channel.
This has the advantage that our equations for $g_0^0$ and $g_1^1$ are independent 
from one another. In contrast, in \cite{Buettiker:2003pp}
they use more subtractions and the subtraction constants are constrained by means of
sum rules that mix the dispersive representations of both waves.

\subsection{Hyperbolic Dispersion Relations}

For their derivation we basically follow the 
same steps described in \cite{Johannesson:1976qp} but using $a\neq0$,
or more recently the steps in \cite{Ditsche:2012fv} 
but applied here to for $\pi \pi \rightarrow K \bar{K}$
instead of $\pi N$ scattering. 
Recall that in this work we use hyperbolas $(s-a)(u-a)=b$, which with $s+t+u=2\Sigma$,
implies that $s$ and $u$ on these hyperbolas are the following functions of $t$:
\begin{eqnarray}
s_b\equiv s_b(t)&= \frac{1}{2}\left(2\Sigma -t+\sqrt{(t+2a-2\Sigma)^2-4b} \right)&, \nonumber \\
u_b\equiv u_b(t)&= \frac{1}{2}\left(2\Sigma -t-\sqrt{(t+2a-2\Sigma)^2-4b} \right)&
\end{eqnarray}

Let us remark that we do not need any subtraction for the antisymmetric amplitude
\begin{align}
&\frac{F^-(s_b,t)}{s_b-u_b}=\frac{1}{2\pi}\int^{\infty}_{4m_{\pi}^2}dt'\frac{\im G^1(t',s'_b)}{(t'-t)(s'_b-u'_b)} \nonumber \\
&\hspace{2cm}+\frac{1}{\pi}\int^{\infty}_{m_+^2}ds'\frac{\im F^-(s',t'_b)}{(s'-s_b)(s'-u_b)},
\label{eq:ahdr}
\end{align}
where
\begin{align}
s'_b&\equiv s_b(t'), \quad u'_b \equiv u_b(t'),\nonumber \\
t'_b&=2\Sigma-s'-\frac{b}{s'-a}+a.
\end{align}
Whereas for the symmetric one:
\begin{align}
& F^+(t,b,a)=h(b,a)+\frac{t}{\pi}\int^{\infty}_{4m_{\pi}^2}\frac{\im G^0(t',s'_b)}{\sqrt{6}\,t'(t'-t)}dt' \nonumber  \\
&+\frac{1}{\pi}\int^{\infty}_{m_+^2}ds'
\frac{\im F^+(s',t'_{b})}{s'}\Big(\frac{s}{s'-s}+\frac{u}{s'-u}\Big).
\label{eq:preshdr}
\end{align}
With these numbers of subtractions the convergence is fast enough so that the asymptotic
amplitude contribution is relatively small (recall it starts at $t=4\,\gev^2$ and $s\simeq 3\,\gev^2$ in this work).
In the above equations $s_b$ and $u_b$ are the values of $s$ and $u$ 
that lie in the hyperbola $(s-a)(u-a)=b$ for a given value of $t$.
Now, we want to rewrite 
the subtraction constant $h(b,a)$ and for this we
follow the procedure in \cite{Johannesson:1976qp,Ananthanarayan:2001uy}.
We thus introduce the following fixed-$t$ dispersion relation
\begin{equation}
F^+(s,t)=c(t) \label{eq:fixtplus}
+\frac{1}{\pi}\int^{\infty}_{m_+^2}ds'\im \frac{F^+(s',t)}{s'^2}\Big(\frac{s^2}{s'-s}+\frac{u^2}{s'-u}\Big).
\end{equation}
Note that two subtractions are needed to ensure the convergence of this 
fixed-t dispersion relation, due to the Pomeron contribution. 
Next, recall that $G^0(t,s,u)=\sqrt{6}F^+(s,t,u)$, so that
by equating Eq.\eqref{eq:preshdr} and \eqref{eq:fixtplus} at $t=0$, $b=a^2-2\Sigma a+\Delta^2$,
the values of $c(t)$ and $h(b,a)$ are determined. Actually,  Eq.\eqref{eq:preshdr}
can be rewritten as:
\begin{align}
&F^+(s_b,t)=8\pi m_+ a_0^++\frac{t}{\pi}\int^{\infty}_{4m_{\pi}^2}\frac{\im G^0(t',s'_b)}{\sqrt{6}\,t'(t'-t)}dt' \nonumber \\
&+\frac{1}{\pi}\int^{\infty}_{m_+^2}ds'\frac{\im F^+(s',t_b)}{s'}\left[h(s',t,b,a)-h(s',0,b,a) \right]\nonumber \\
&+\frac{1}{\pi}\int^{\infty}_{m_+^2}ds'\frac{\im F^+(s',0)}{s'^2} \left[ g(s',b,a)-g(s',\Delta^2,0) \right], 
\label{eq:shdr}
\end{align}
where
\begin{align}
&h(s',t,b,a)=\frac{s'(2\Sigma-t)-2[b-a^2+(2\Sigma-t)a]}{s'^2-s'(2\Sigma-t)+[b-a^2+(2\Sigma-t)a]},\nonumber\\
& g(s',b,a)=\frac{s'(2\Sigma)^2-2[b-a^2+2\Sigma a](s'+\Sigma)}{s'^2-s'2\Sigma+[b-a^2+2\Sigma a]}.
\end{align}
We have explicitly checked that in the  $a=0$ case we recover the
HDR in \cite{Johannesson:1974ma, Johannesson:1976qp, Ananthanarayan:2001uy}. 
However, with our HDR above we can now choose the $a$ parameter to maximize
the applicability region of the HDR once projected into partial waves, which we will do
in the next subsection.

Before finishing this subsection, a comment on the high energy region is in order.
We have three different kinds of
contributions above 2 GeV, the first one is $G^I(t',s'_b)$, which 
can be calculated from Eq. \eqref{eq:vene}. The second kind is the evaluation of $F^{\pm}(s',0)$:
for the symmetric amplitude we just use Eq.\eqref{eq:reggef+}, while for the anti-symmetric one 
we use Eq.\eqref{eq:reggef-}. The last kind is  for $F^\pm(s',t'_b)$, 
which corresponds to an exotic exchange, so that its contribution is negligible.

\subsection{Partial-wave hyperbolic dispersion relations}

In this work we want to obtain parameterizations of the
 $\ell=0,1,2$ partial waves which are consistent with data and the hyperbolic dispersive representation.
Thus, we project Eqs.\eqref{eq:ahdr} and \eqref{eq:shdr} into partial waves using Eq.\eqref{proj} 
 to obtain a set of Roy-Steiner-like equations:
\begin{widetext}
\begin{align}
&g^0_0(t)=\frac{\sqrt{3}}{2}m_+a^+_0+\frac{t}{\pi}\int^{\infty}_{4m_{\pi}^2}\frac{\im g^0_0(t')}{t'(t'-t)}dt' +\frac{t}{\pi}\sum_{\ell\geq2}\int^{\infty}_{4m_{\pi}^2}\frac{dt'}{t'} G^0_{0, 2\ell-2}(t,t') \im g^0_{2\ell-2}(t') +\frac{1}{\pi}\sum_\ell \int^{\infty}_{m_+^2}ds' G^+_{0, \ell}(t,s') \im f^+_\ell(s'),\nonumber \\
&g^1_1(t)=\frac{1}{\pi}\int^{\infty}_{4m_{\pi}^2}\frac{\im g^1_1(t')}{t'-t}dt' +\frac{1}{\pi}\sum_{\ell\geq2}\int^{\infty}_{4m_{\pi}^2}dt' G^1_{1, 2\ell-1}(t,t') \im g^1_{2\ell-1}(t') +\frac{1}{\pi}\sum_\ell \int^{\infty}_{m_+^2}ds' G^-_{1, \ell}(t,s') \im f^-_\ell(s'), \nonumber \\
&g^0_2(t)=\frac{t}{\pi}\int^{\infty}_{4m_{\pi}^2}\frac{\im g^0_2(t')}{t'(t'-t)}dt' +\frac{t}{\pi}\sum_{\ell\geq2}\int^{\infty}_{4m_{\pi}^2}\frac{dt'}{t'} G'^0_{2, 4\ell-2}(t,t') \im g^0_{4\ell-2}(t') +\frac{1}{\pi}\sum_\ell \int^{\infty}_{m_+^2}ds' G'^+_{2, \ell}(t,s') \im f^+_\ell(s').
\label{eq:pwhdr}
\end{align}
\end{widetext}

The explicit expressions of the $G^I_{\ell \ell'}(t,t'),G^\pm_{\ell \ell'}(t,s')$ integration kernels are given in Appendix~\ref{app:kernels}. Since so far in this work we have left free the $a$ parameter, we can now use it to maximize the
applicability of the equations right above. Note there are constraints coming from the applicability of the HDR
in Eqs.\eqref{eq:ahdr} and \eqref{eq:shdr} as well as from the convergence of the partial-wave expansion.
As shown in appendix \ref{app:convergence}, by setting $a=-10.8m_{\pi}^2$ 
the applicability range of these equations 
is $-0.286 \,\gev^2\leq t\leq 2.19 \,\gev^2$. 
In other words, we can study the physical region from the $K \bar K$ threshold $\simeq 0.992\,\gev$ up to 
$\simeq 1.47\,\gev$.
In contrast, the usual HDR projected into partial waves are only valid up to $\simeq 1.3,\gev$.
Thus, with our choice of $a$, 
the applicability of the dispersive approach in the physical region, where we can test or use data as input, has been extended by 55\% in terms of the $\sqrt{t}$ variable, or 67\% in terms of $t$.

As can be directly seen in Eq.\eqref{eq:pwhdr} the $g^1_1(t)$ partial wave does not have any scattering length as input parameter and its dominant contribution to the integral comes
from its own imaginary part. Since it is not subtracted, the Regge contribution is not negligible, but we have already attached a conservatively large uncertainty to its residue and we will see that it barely changes when using the dispersive representation as a constraint on data. In the case of even partial waves, one subtraction is necessary to ensure the convergence, 
and hence the output is always influenced by the scattering lengths coming from $\pi K $ scattering. 
In this work we fix them to the values obtained in \cite{Pelaez:2016tgi}, 
which are also compatible with the Roy-Steiner prediction in \cite{Buettiker:2003pp}.
As already commented, an important advantage of using HDR with the smaller possible number 
of subtractions is to decouple odd and even partial waves. For example in \cite{Buettiker:2003pp} the 
Roy-Steiner equation for $g^0_0$ uses $g^1_1$ as input. 

Finally, we want to remark that, as usual, the high energy part of the integrals in Eqs.\eqref{eq:pwhdr}
is obtained by projecting into the corresponding partial-wave the high-energy part of the integrals in Eqs.\eqref{eq:ahdr} and \eqref{eq:shdr}, where Regge theory was used as input as explained in previous sections.

\subsection{The unphysical region and the Muskhelishvili-Omn\`es problem}

As can be observed in Eqs.\eqref{eq:pwhdr}, the integration region
actually starts at $\pi \pi$ threshold. This means that the integrals extend over an
``unphysical'' regime where $\pi\pi\rightarrow K \bar K$ scattering does not occur and 
thus cannot be described with data parameterizations.
Nevertheless,  below $K\bar K$ threshold the inelasticity to more than two-pion states 
is completely negligible. Since $\pi\pi$ is the only available state in that region 
 Watson's Theorem implies that the $g_\ell^{I_t}$ phase 
below $K\bar K$ threshold
is just that of $\pi \pi$ scattering and thus we write $\phi^{I_t}_\ell(t)=\delta^{I_t}_{\ell,\pi \pi \rightarrow \pi \pi}(t)$. Note that Watson's Theorem does not provide any direct information on $\vert g_\ell^{I_t}\vert$.
But once the phase is known, determining the 
modulus in the unphysical region is nothing but the standard  
Muskhelishvili-Omn\`es problem \cite{Omnes}, that we describe next
following similar steps as in \cite{Johannesson:1976qp,Ananthanarayan:2001uy,Buettiker:2003pp,Hoferichter:2011wk,Ditsche:2012fv}. 
Recalling that partial waves have a right- and left-hand cut
we can re-write Eqs.\eqref{eq:pwhdr} as follows:
\begin{align}
&g^0_\ell(t)=\Delta^0_\ell(t) + \frac{t}{\pi}\int^{\infty}_{4m_\pi^2}\frac{dt'}{t'}\frac{\im g^0_\ell(t)}{t'-t}, \quad \ell=0,2,\nonumber \\
&g^1_1(t)=\Delta^1_1(t) + \frac{1}{\pi}\int^{\infty}_{4m_\pi^2}dt'\frac{\im g^1_1(t)}{t'-t}, 
\end{align}
where the $\Delta^{I}_\ell(t)$ contain the left-hand cut contributions and subtraction terms.
Note that $\Delta^{I}_\ell(t)$ does not depend on $g^I_\ell$ itself, but on 
other $g^I_{\ell'}$ with $\ell'\geq\ell+2$,
which in the unphysical region are much more suppressed than $g^I_\ell$, due to the centrifugal barrier.

Now we define the Omn\`es function
\begin{equation}
\Omega^I_\ell(t)=\exp\left(\frac{t}{\pi}\int^{t_m}_{4m_\pi^2}\frac{\phi^I_\ell(t')dt'}{t'(t'-t)}\right),
\end{equation}
which satisfies 
\begin{equation}
\Omega^I_\ell(t)\equiv\Omega^I_{l,R}(t)e^{i\phi^I_\ell(t)\theta(t-4m_\pi^2) \theta(t_m-t)}, 
\end{equation}
where, in the real axis, $\Omega^I_{l,R}(t)$ can be written as:
\begin{eqnarray}
\Omega^I_{l,R}(t)&=&\left\vert \frac{t_m}{t_\pi}(t-t_\pi)^{-\phi^I_\ell(t)/\pi}(t_m-t)^{\phi^I_\ell(t)/\pi}\right\vert 
\nonumber\\ &&
\times \exp\left(\frac{t}{\pi}\int^{t_m}_{4m_\pi^2}dt'\frac{\phi^I_\ell(t')-\phi^I_\ell(t)}{t'(t'-t)}\right).
\end{eqnarray} 
In the real axis, $\Omega^I_{l,R}$ is nothing but the modulus of $\Omega^I_{l}$ 
and therefore a real function.

Note that from $4m_\pi^2$ to $t_m$ the Omn\'es function has the same cut 
as $g^I_\ell(t)$. Thus, we can define a function
\begin{align}
F^I_\ell(t)=\frac{g^I_\ell(t)-\Delta^I_\ell(t)}{\Omega^I_\ell(t)},
\end{align}
which is analytic except for a right hand cut starting at $t_m$. 
Hence we can write dispersion relations 
for $F^I_\ell(t)$, which in terms of  $g^I_\ell(t)$ read:
\begin{eqnarray}
g^0_0(t)&=&\Delta^0_0(t)+\frac{t\Omega^0_0(t)}{t_m-t}\left[\raisebox{0pt}[0.6cm][0pt]{} \alpha \right. \nonumber \\
&&\left.+\frac{t}{\pi}\int^{t_m}_{4m_\pi^2}dt'\frac{(t_m-t')\Delta^0_0(t')\sin\phi^0_0(t')}{\Omega^0_{0,R}(t')t'^2(t'-t)}\right. \nonumber \\
&&\left.+\frac{t}{\pi}\int^{\infty}_{t_m}dt'\frac{(t_m-t')\vert g^0_0(t')\vert \sin\phi^0_0(t')}{\Omega^0_{0,R}(t')t'^2(t'-t)} \right],
\label{MO00}\\
g^1_1(t)&=&\Delta^1_1(t)+\Omega^1_1(t)\left[\frac{1}{\pi}\int^{t_m}_{4m_\pi^2}dt'\frac{\Delta^1_1(t')\sin \phi^1_1(t')}{\Omega^1_{1,R}(t')(t'-t)} \right. \nonumber \\
&&\left.+\frac{1}{\pi}\int^{\infty}_{t_m}dt' \frac{\vert g^1_1(t')\vert \sin \phi^1_1(t')}{\Omega^1_{1,R}(t')(t'-t)}\right], 
\label{MO11}\\
g^0_2(t)&=&\Delta^0_2(t)+t\Omega^0_2(t)\left[\frac{1}{\pi}\int^{t_m}_{4m_\pi^2}dt'\frac{\Delta^0_2(t')\sin \phi^0_2(t')}{\Omega^0_{2,R}(t')
t'(t'-t)} \right. \nonumber \\
&&\left.+\frac{1}{\pi}\int^{\infty}_{t_m}dt' \frac{\vert g^0_2(t')\vert \sin \phi^0_2(t')}{\Omega^0_{2,R}(t')t'(t'-t)}\right].
\label{MO20}
\end{eqnarray} 
When $t$ lies in the real axis above the $\pi\pi$ threshold, 
a principal value must be understood on each integral.
In addition, between $\pi\pi$ threshold and $t_m$ 
on the left hand sides the amplitude is reduced to its modulus 
(since by construction the Omn\'es function removes the phase), whereas above $t_m$  it is reduced to its real part.

Since in the next sections we will choose $t_m$ with $\phi^0_0(t_m)\geq \pi$ we have introduced one subtraction for the $g^0_0(t)$ Omn\`es solution in order to ensure the convergence when $t\rightarrow t_m$. The subtraction constant $\alpha$ will be obtained by imposing numerically 
a no-cusp condition on $t_m$ for $g^0_0(t)$.

The interest of these equations is that for a given $g^I_\ell(t)$, the integrals
{\it in the unphysical region} only make use of the phases
and the $\Delta^I_\ell$. But thanks to Watson's Theorem the former are known from 
$\pi\pi$ scattering, which we take from the dispersive analysis of \cite{GarciaMartin:2011cn},
and the latter do not involve $g^I_\ell(t)$ itself, but only 
partial waves with $\ell'-\ell\geq 2$.
These higher partial waves are suppressed in the unphysical 
region with respect to that with $\ell$.
 We also need input from $K \pi$ scattering that is known and we take it
from our recent dispersive data analysis in \cite{Pelaez:2016tgi}.
Thus we can directly solve $g^1_1(t)$ and $g^0_2(t)$, for which we have explicitly 
checked that the $\ell=3$ and $\ell=4$ contributions are small and negligible, respectively. 
Once we have $g^0_2(t)$ we can use it as input 
to solve Eq.\eqref{MO00} for $g^0_0(t)$.

It is worth noticing here that, in purity, for the Regge contributions to $\Delta^I_\ell(t)$, 
one has to subtract the projection of the 
Regge amplitude itself into the desired $I,\ell$ partial wave. 
Fortunately this projection is negligible, 
and our solutions do not depend on this procedure.

We still have to discuss the choice of $t_m$, which is always above the $K \bar K$ threshold.
It is important to recall that the derivation
of the above equations implies that $g_{output}(t_m)=g_{input}(t_m)$.
This condition will always be forced into the output no matter if the data at that energy is in good or bad agreement with dispersion relations.
If the data at that energy region were not close to the dispersive solution, the output 
would be forced to describe it and the result could be strongly distorted in other regions. 
In particular the $g^0_0$ wave is the most sensitive to this instability, the effect is more moderate on the $g^0_2$ 
and negligible for the $g^1_1$ because it is already very consistent for any $t_m$ choice.
Thus, we have studied what energy region is the most consistent for $g^0_0$ when changing $t_m$ 
and we have found that there are two regions that yield systematically rather consistent results between input and output: one around $\sqrt{t_m}=1.2\, \gev$, which is also valid for $g_2^0$,
and another one around $\sqrt{t_m}=1.47\, \gev$.
However, if we chose the latter, we find that the uncertainty in the dispersive result between between $K \bar K$
and 1.2 GeV is so large that there is no dispersive constraint in practice, having larger uncertainties could even produce both $g^0_0(t)$ solutions to be compatible between them. Moreover 
by looking at Eqs.~\eqref{MO00},\eqref{MO11} and \eqref{MO20} one can notice that $t_m$
marks the energy above which $\vert g_\ell^I\vert $ is used as input for its own equation. Since we are actually
trying to test the data parameterizations, within our approach we would like to maximize that region and choose the smaller possible $t_m$.
All in all, we have made the final choice $\sqrt{t_m}=1.2\, \gev$ for all partial waves. This is a point above $K\bar K$ threshold 
where there are no cusps coming from the two most important inelasticities 
($K\bar K$, $\eta \eta$). In particular, the $g^0_2$ is well controlled 
at this energy since its largest 
contribution comes from the $f_2(1270)$, a very well-known resonance very close to $t_m$.

\section{Consistency check of unconstrained fits}
\label{sec:Consistency}



In order to study in a systematic way the consistency of 
the unconstrained data parameterizations of Sect.\ref{sec:UFD} with respect to
dispersion relations, we first define a ``distance-square'' 
\begin{equation} 
d^2=\frac{1}{N}\sum_{i=1}^{N} \left(\frac{d_i}{\Delta d_i}\right)^2,
\label{eq:distances}
\end{equation}
for each dispersion relation. 
Note its similarity to a $\chi^2$ function, although we are still not fitting or imposing the dispersion relations. Here $d_i$ is
the difference between the ``input'' and ``output'' of each dispersion relation at
the energy $\sqrt{t_i}$. 
We use thirty energy points $\sqrt{t_i}$ equally spaced 
from threshold up to 1.47~GeV.
In addition, $\Delta d_i$ is the uncertainty in the $d_i$ 
difference, which is obtained by varying the parameters of our
unconstrained fits to data (UFD) within their errors. 

As we explained before,  Eqs.\eqref{MO00},\eqref{MO11},\eqref{MO20} yield the modulus of 
the partial wave below $t_m$ and the real part above. 
However, in order to simplify our plots and calculations,
we will just display the modulus. In particular
by ``input" we will understand the modulus of the partial wave on
the left hand side of  Eqs.\eqref{MO00},\eqref{MO11},\eqref{MO20}, i.e. 
as obtained directly from our fits. Similarly, by ``output" we
will always mean the modulus of the dispersive representation.
Note that for $t<t_m$ 
this modulus is obtained from the right hand side of those equations 
with principal values on each integral.
However, for $t>t_m$ only the real part is obtained from the integrals
and the modulus is reconstructed by adding the imaginary
part from the direct parameterizations.

With the above definition we can study the consistency of each partial-wave dispersion relation.
It will be well satisfied on the average if its corresponding
$d^2\leq 1$. In case of disagreement it is also 
relevant to check whether it comes from a particular energy region and for this we will show
figures comparing the input and output as a function of $\sqrt{t}$.

\subsection{$g^1_1$ UFD check}

Let us study first the consistency of $g^1_1$. We see in  Eq.\eqref{MO11} that its partial wave dispersion relation is decoupled from even partial waves.  
The highest partial wave we have considered in $\Delta^1_1$ is the $\ell=3$ contribution. Actually, by using the simple model dominated by the $\rho(1690)$ resonance described in Sect.\ref{sec:higherwaves}, we have explicitly checked that its contribution is very small and barely affects our results for $g^1_1$ below 1.47 GeV.

\begin{figure}
\centering
\includegraphics[scale=0.32]{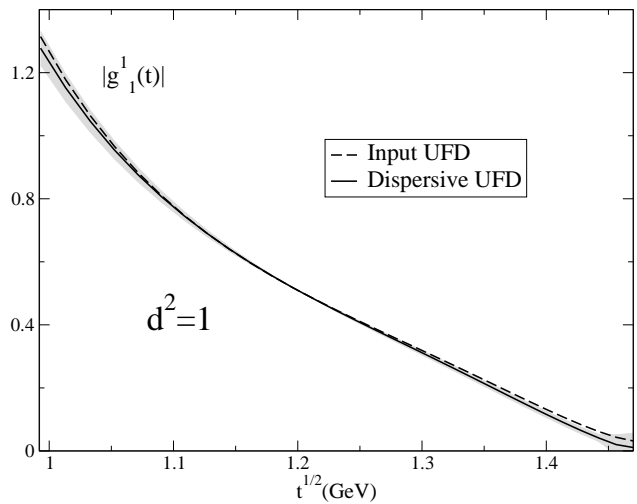}
\caption{\label{fig:checkg11} Comparison between the input (dashed line) and the dispersive output (continuous line) for the modulus of the $g^1_1$ dispersion relation in Eq.\eqref{MO11}. The gray band covers the uncertainty of the difference.
 }
\end{figure}

As can bee seen in Fig.\ref{fig:checkg11} the dispersion relation in Eq.\eqref{MO11} is remarkably well satisfied, with a total $d^2=1$. 
Such a nice agreement was expected since it has a large contribution
from  the $\rho(770)$ that dominates 
 $\pi \pi$ scattering in this channel below $K \bar K$ threshold,  and our input from \cite{Pelaez:2016tgi} is already consistent with $\pi\pi$ data and dispersion relations.
Let us now recall that the $\pi\pi\rightarrow K \bar K$ data we use as input show
large uncertainties and fluctuations (see Fig.\ref{fig:g11ufd}). Our UFD description does not follow visually
all these fluctuations but, roughly speaking, it averages them and rises softly and monotonously.
Still, our UFD is remarkably consistent with the dispersive representation. Actually we have checked that parameterizations with more oscillations may describe the central values of the data points better, but satisfy worse the dispersive representation than  
our UFD fit. In the $\pi\pi\rightarrow K \bar K$ physical region
we had also included resonant shapes for the $\rho'$ and $\rho''$ resonances in our UFD.
As seen from our results, the parameters and shape of the $\rho'$, 
which for a good part lies within the applicability region of our 
equations, are fairly consistent with dispersion relations. As commented in Sect.\ref{sec:UFD11} the $\rho''$ was used just
as a simple form to parameterize the amplitude at energies beyond the reach of our dispersive representation where scattering data do not exist. 

One could also be worried that, since the $g^1_1$ dispersion relation has no
subtractions, it may require some tuning on the Regge asymptotics and the $\lambda$ parameter we estimated with the Veneziano model and degeneracy in subsection \ref{sec:Regge}. However the nice fulfillment of the dispersion relation yields strong support for our $\lambda$ estimations.

\subsection{$g^0_2$ UFD check}

\begin{figure}
\centering
\includegraphics[scale=0.32]{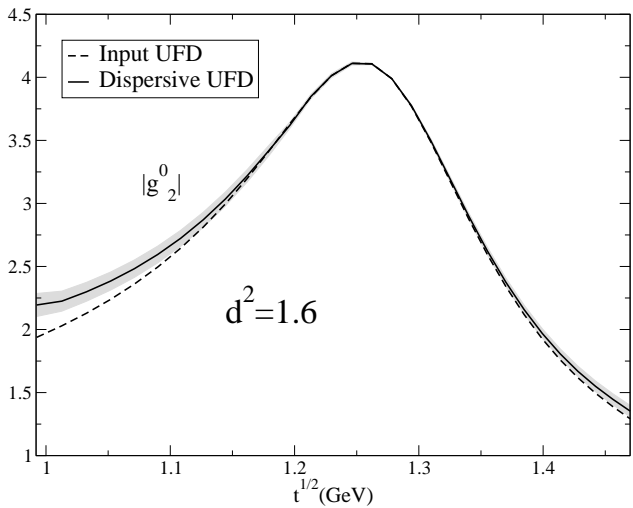}\vspace*{-.3cm}
\caption{ \label{fig:checkg02} Comparison between the input (dashed line) and the dispersive output (continuous line) for the modulus of the $g^0_2$ dispersion relation in Eq.\eqref{MO20} using as input the UFD set. The gray band covers the uncertainty of the difference.}
\end{figure}

In the case of the $g^0_2(t)$ dispersion relation, Eq.\eqref{MO20}, it involves even partial waves with $\ell\geq4$,
but they are almost negligible below 2 GeV. As seen in Fig.~\ref{fig:checkg02}, when using the UFD parameterizations,
the $g^0_2(t)$ dispersion relation is clearly not well satisfied right above $K\bar K$ threshold and this incompatibility
fades away near 1.1 GeV. At threshold, the deviation is $\simeq3\sigma$. Very naively one could have expected this region to be dominated by the $f_2(1270)$ resonance tail, since the threshold is merely 1.5 widths away from the resonance peak. However, if one tries to use a simple Breit-Wigner description instead of our UFD parameterization, then $d^2\geq 6$. Thus, such naive expectation does not hold, which justifies the elaborated form of our parameterization in Eq.\ref{eq:g02}. Nevertheless, there is still room for improvement that will be achieved when imposing the dispersion relations as constraints in Section~\ref{sec:CFD}.

\begin{figure}
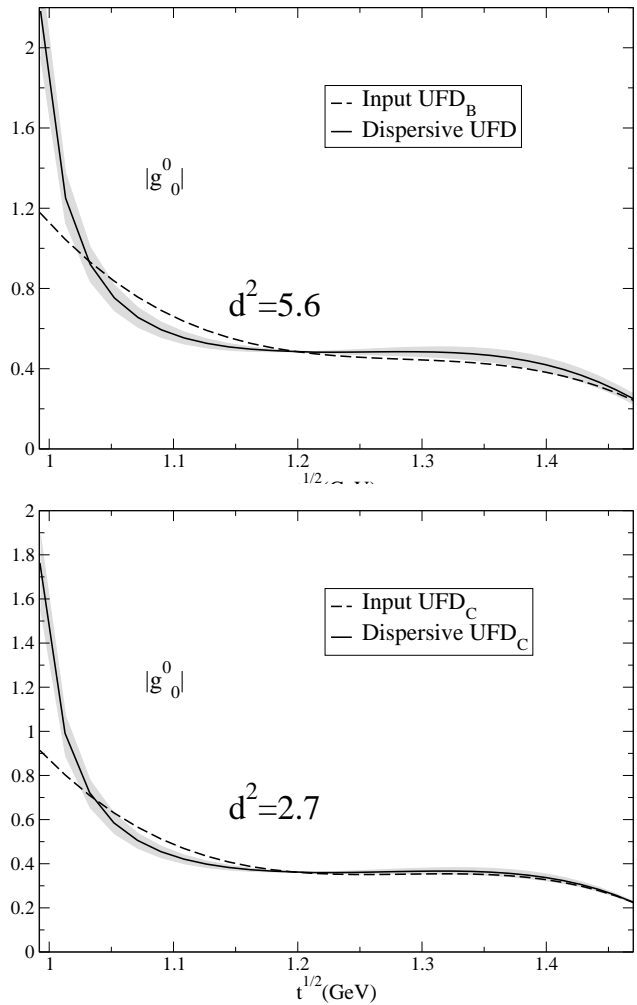

\centering
\includegraphics[scale=0.32]{checkg00.eps}

\includegraphics[scale=0.32]{checkg00cohen.eps}\vspace*{-.3cm}
\caption{ \label{fig:checkg00}  Comparison between the input (dashed line) and the dispersive output (continuous line) for the modulus of the $g^0_0$ dispersion relation in Eq.\eqref{MO00}. In the upper panel we show the results using as input the UFD$_\text{B}$ parameterization and in the lower panel those from the UFD$_\text{C}$. The gray bands cover the uncertainty of the difference between the input and the respective dispersive result.
}
\end{figure}

\subsection{$g^0_0$ UFD check}

Finally, for the scalar-isoscalar dispersion relation in Eq.\eqref{MO00}, we need both the $g^0_0(t)$ and $g^0_2(t)$. In this case, partial waves with $\ell\geq 4$ are totally negligible below 2 GeV. In Fig.\ref{fig:checkg00} we show the results of the $g^0_0(t)$ dispersion relation when using either the UFD$_\text{B}$ or UFD$_\text{C}$ parameterizations as input.
In both cases the agreement is poor, particularly due to the results in the region 10-20 MeV above $K\bar K$ threshold, where the dispersive solution increases rapidly. 
This feature is common to both the UFD$_\text{B}$ and UFD$_\text{C}$ and is due to the influence of the $f_0(980)$.  The respective $d^2=5.6$ and $d^2=2.7$ are dominated by this near threshold region. There is a clear need for improvement, that we will achieve by imposing dispersion relations as constraints in the next section, although in both cases the disagreement in the region very near threshold will linger on. However, we will see that for both solutions a very good consistency with dispersion relations can be achieved except for the very near threshold region.

Finally, let us remark that the $g_0^0$ partial-wave dispersion relation in Eq.\eqref{MO00} depends on the $\pi K$ scattering length $a^+_0$. We have checked that the dispersion relation would be better satisfied
if we used a somewhat lower value of $a^+_0$ than that obtained in our previous work \cite{Pelaez:2016tgi} (which was also compatible with Roy-Steiner determinations \cite{Buettiker:2003pp}). Since in this work we are considering $\pi K$ scattering 
amplitudes as fixed input, we keep the value from the $\pi K$ constrained fit, but this result could be relevant for future re-analysis of $\pi K$ scattering data.

\section{Constrained Fits to Data}
\label{sec:CFD}

Therefore, we have just seen that the data on the $g_2^0$ and even more so on the $g_0^0$ do not satisfy very well the dispersive representation. There is clear room for improvement. Thus, in this section we will impose the dispersion relations in Eqs.\eqref{MO00}, \eqref{MO11}, \eqref{MO20} as constraints of the fits. In this way we will obtain a set of Constrained Fits to Data (CFD) which fulfillment of the dispersive representation will be much improved. In this section we use the same functional forms for the amplitudes that we used in Sect.\ref{sec:UFD}, but the parameters
change from the UFD to the CFD sets. In general the difference between the UFD and CFD parameters is small, with a few exceptions. Nevertheless, due to large correlations in the parameters, even if some CFD parameters deviate from the UFD set, the resulting UFD and CFD curves
are typically consistent with one another at the 1 or 1.5 $\sigma$ level. Only for the constrained analysis of the UFD$_\text{C}$, the CFD$_\text{C}$ $g_0^0$ partial wave deviates by about 2 $\sigma$ in the region from 1.25 to 1.45 GeV, but it still compatible with the upper error bars of the data.  Hence the CFD description of data is still rather good.


To minimize the discrepancy between the fit used as input in the dispersion relation 
and the output obtained from the dispersion relation, without deviating much from the data,
one first defines a $\chi^2$-like function
\begin{align}
W_1^2&d_{g^I_\ell}^2+\frac{W_2^2}{N}\sum^N_k \left (\frac{|g^I_\ell|_{exp,k}-|g^I_\ell(s_k)|}{\delta |g^I_\ell|_{exp,k}}\right)^2  \nonumber\\
&+\frac{W_3^2}{N'}\sum^{N'}_k \left (\frac{(\phi^I_\ell)_{exp,k}-\phi^I_\ell(s_k)}{\delta (\phi^I_\ell)_{exp,k}}\right)^2,
\label{eq:weminimize}
\end{align}
where $|g^I_\ell|_{exp,k},(\phi^I_\ell)_{exp,k}$ are the experimental values of the $k$-th data point for the modulus and the phase, respectively, and  $\delta|g^I_\ell|_{exp,k},\delta(\phi^I_\ell)_{exp,k}$ are their corresponding errors. The weights $W_1^2$, $W^2_2=W'^2N/(N+N')$, $W^2_3=W'^2N'/(N+N')$ are used to 
roughly take into account the degrees of freedom needed to parameterize the curves that describe the modulus and the phase.  For simplicity we have chosen the same $W_1^2=5$ and $W'^2=12$ value for all partial waves as an average value of their degrees of freedom.
Note that we actually minimize the sum of this function over the three partial waves of interest $(I,\ell)=(0,0),(1,1)$ and $(0,2)$. In addition, recall that, as explained in Sec.~\ref{subsec:ufd20}, we have added two points to the $\chi^2$-function to take into account the experimental mass of the $f_2$ and $f_2'$ resonances.

Let us remark that in previous works our procedure was slightly different:
we defined a similar $\chi^2$-like function but in terms of the unconstrained fit parameters, which were not allowed to vary much from their unconstrained best values. In contrast, in Eq.~\eqref{eq:weminimize} we define our $\chi^2$-like function
directly in terms of data, not the unconstrained fit parameters. 
The reason is that in this work the onset of Regge parameterizations is 2 GeV and 
thus we use our partial-wave parameterizations to describe data from $K \bar K$ threshold up to 2 GeV. However, the dispersion relations are only applicable up to 1.47 GeV. If we constrained only the fit parameters with the dispersion relations, which affect only the lower-energy data, we would obtain large artificial deviations in the description of the higher-energy data. 
With the procedure we use here, and contrary to what happened in previous works, if there are some strongly correlated parameters,
we can see that their constrained values can deviate appreciably from their unconstrained best values but still the constrained and unconstrained curves look very similar. As the uncertainty variation is of second order, and parameters that are not compatible with old values deviate by a small number of sigmas at most, we still maintain their uncertainties as they are a reliable and almost unchanged estimate of the error, as one can see in the final uncertainty band plotted in the figures for the CFD parameterizations.

\subsection{Constrained $g^ 1_1(t)$ partial wave}

\begin{figure}
\centering
\includegraphics[scale=0.32]{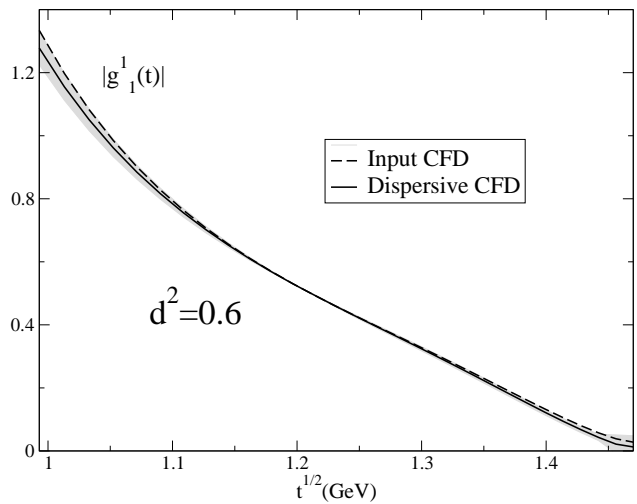}
\caption{ Comparison of the modulus and the dispersion relation after the minimization procedure. The gray band covers the uncertainty of the difference between the input and dispersive results.
 \label{fig:cfd11}}
\end{figure}

Let us recall that the UFD $I=1,\ell=1$ wave from $K\bar K$ threshold up to 1.47 was already consistent with the dispersive representation. By imposing our dispersion relations $d^2$ decreases just from 1 to 0.6. The difference between the constrained input and dispersive output for the $g^1_1$ wave can be seen in Fig.~\ref{fig:cfd11}.

\begin{figure}
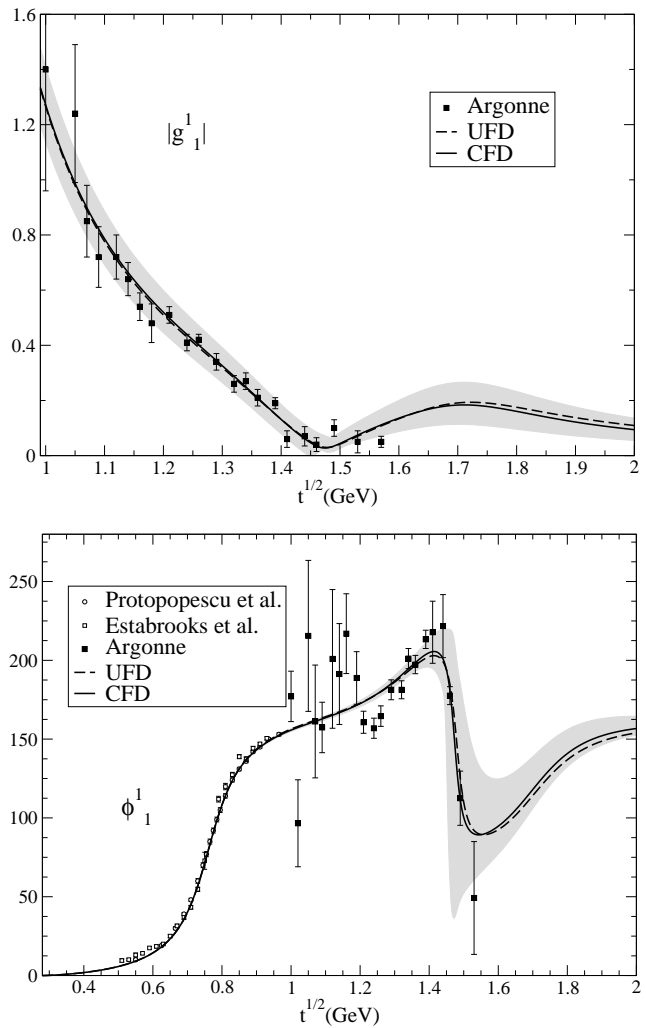

\centering
\includegraphics[scale=0.32]{amplig11cfd.eps}\vspace*{.3cm}
\includegraphics[scale=0.32]{phaseg11cfd.eps}
\caption{\rm \label{fig:g11cfd} 
Modulus and phase of the $g_1^1(t)$ $\pi\pi\rightarrow K \bar K$ partial wave. 
The continuous line 
and the uncertainty band correspond to the CFD 
parameterization, whereas the dashed line corresponds to the UFD. 
The white circles and squares come from the $\pi\pi$ scattering experiments of
Protopopescu et al. 
\cite{Protopopescu:1973sh} and Estabrooks et al.\cite{Estabrooks:1974vu}, respectively.}
\end{figure}

\begin{figure}
\centering
\includegraphics[scale=0.32]{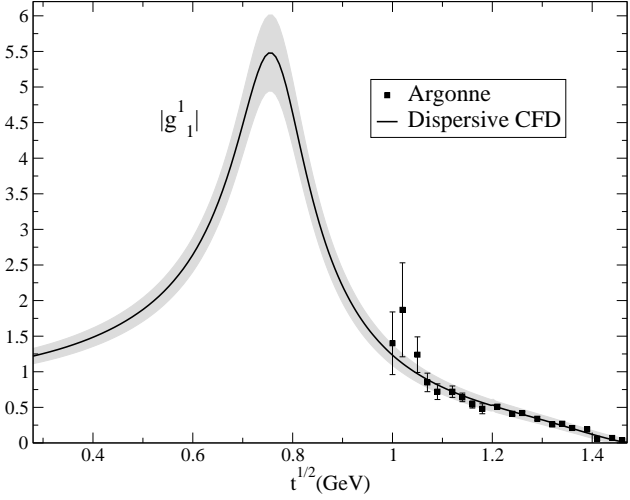}\vspace*{.3cm}
\caption{\rm \label{fig:g11cfddispersive} 
Dispersive output for the modulus of the $g_1^1(t)$ $\pi\pi\rightarrow K \bar K$ partial wave obtained from the CFD set. 
The continuous line 
and the uncertainty band corresponds to the CFD dispersive result.}
\end{figure}

Actually, as seen in in Fig.~\ref{fig:g11cfd} imposing the dispersive constraints barely changes this wave, i.e. the UFD and CFD curves are almost indistinguishable both for the modulus and the phase of $g_1^1$.  Note also that, as shown in Fig.~\ref{fig:g11cfddispersive}, the dispersive CFD output perfectly describes the data. In that Figure we also show the CFD modulus in the unphysical region and the continuous matching at threshold.

The new CFD parameters can be found in Table~\ref{tab:g11para} where it can be checked that the CFD values are remarkably consistent with the UFD ones: only two are beyond one standard deviation but not more than 2$\sigma$. As we are using a non-subtracted HDR to study the odd angular momentum partial waves, the small improvement in the description of this partial wave comes mostly from the slight variation of the Regge parameters. Nevertheless, as it can be seen in Table~\ref{tab:reggepiK}, our CFD result for the $\lambda$ Regge parameter is compatible with its UFD value, thus supporting the degeneracy between the $\rho$ and $K^*$ families.

 It is worth noticing that, as we are using no subtractions, the value of the $\pi \pi \rightarrow K \bar K$ amplitude at $t=0$, $b=\Delta^2$ can be related to the $a_0^-$ $\pi K \rightarrow \pi K$ scattering length $a^-_0=(a^{1/2}-a^{3/2})/3$, 
 using Eq.\eqref{eq:ahdr}, to obtain the following sum rule \cite{Karabarbounis:1980bk,Buettiker:2003pp}:
\begin{align}
&\frac{8\pi m_+a^-_0}{m^2_+-m^2_-}=\frac{1}{2\pi}\int^{\infty}_{4m_{\pi}^2}\frac{dt'}{t'}\frac{\im G^1(t',s'_{\Delta^2})}{\sqrt{(t'-4m^2_\pi)(t'-4m^2_K)}} \nonumber \\
&\hspace{2cm}+\frac{1}{\pi}\int^{\infty}_{m_+^2}ds'\frac{\im F^-(s',t'_{\Delta^2})}{\lambda_{s'}}
\end{align}
Note that the scattering length results from the integration over both $\pi K \rightarrow \pi K$ and $\pi \pi \rightarrow K \bar K$ channels. Using as input for $G^1$ our constrained parameterizations just calculated and our
the CFD parameterizations for $K\pi$ scattering in \cite{Pelaez:2016tgi}, we find
\begin{equation}
m_\pi(a^{1/2}-a^{3/2})=0.249\pm0.032, \quad \mbox{(sum rule+CFD)}.
\end{equation}
To be compared with 
\begin{equation}
m_\pi(a^{1/2}-a^{3/2})=0.251\pm0.014, \quad \mbox{(sum rule in \cite{Buettiker:2003pp})}\nonumber 
\end{equation}
obtained in \cite{Buettiker:2003pp} using this same sum rule with their unconstrained input from $\pi\pi\rightarrow K\bar K$ and the $K\pi$ solutions from their Roy-Steiner analysis of $K\pi$. 
We obtain a larger uncertainty since we use the Regge asymptotics from 2 GeV instead of 2.5 GeV as in \cite{Buettiker:2003pp}
and because, in contrast to \cite{Buettiker:2003pp}, we also include uncertainties in all partial-waves.

Those two values obtained using the sum rule can also be compared with direct calculations from the $K\pi$ amplitudes:
\begin{align}
&m_\pi(a^{1/2}-a^{3/2})=0.273^{+0.018}_{-0.015}, \quad \mbox{(CFD \cite{Pelaez:2016tgi})} \nonumber \\
&m_\pi(a^{1/2}-a^{3/2})=0.269^{+0.015}_{-0.015}. \quad \mbox{(Roy-Steiner \cite{Buettiker:2003pp})}\nonumber  
\label{result1}
\end{align} 
The first is obtained from our recent dispersive analysis using Forward Dispersion Relations as constraints on fits to $K\pi$ data \cite{Pelaez:2016tgi} and the second from the solutions of Roy-Steiner equations in \cite{Buettiker:2003pp}.


\begin{figure}
\centering
\includegraphics[scale=0.32]{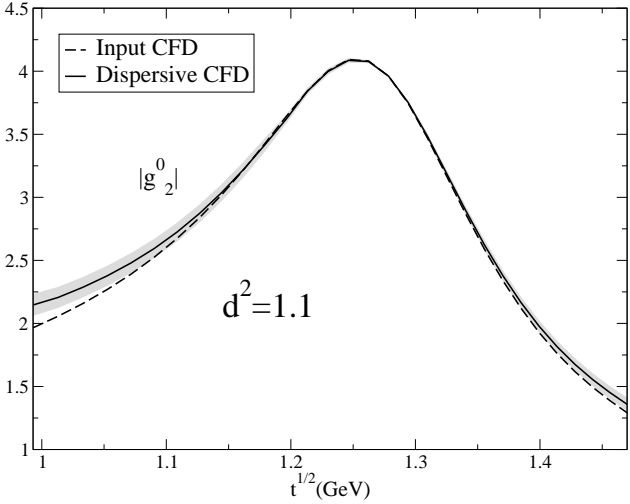}
\caption{Comparison between the input (dashed line) and the dispersive output (continuous line) for the modulus of the $g^0_2$ dispersion relation in Eq.\eqref{MO20} using as input the CFD set. The gray band covers the uncertainty of the difference. \label{fig:cfd02} }
\end{figure}

\begin{figure}
\includegraphics[scale=0.32]{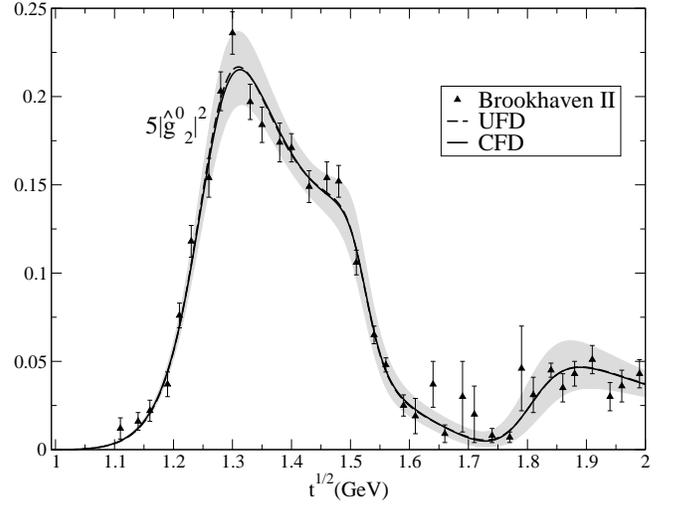}\vspace*{.3cm}
\caption{\rm \label{fig:g02cfd} The continuous line is our final CFD parameterization of the
data on the modulus of $\hat g^0_2(t)$
from the Brookhaven-II analysis \cite{Longacre:1986fh}. The gray band stands for the uncertainty from the CFD parameters.The dashed line is the UFD parameterization. The difference between the UFD and CFD parameterization near threshold is imperceptible due to the $q^5$ factor.
}
\end{figure}

\begin{figure}
\includegraphics[scale=0.32]{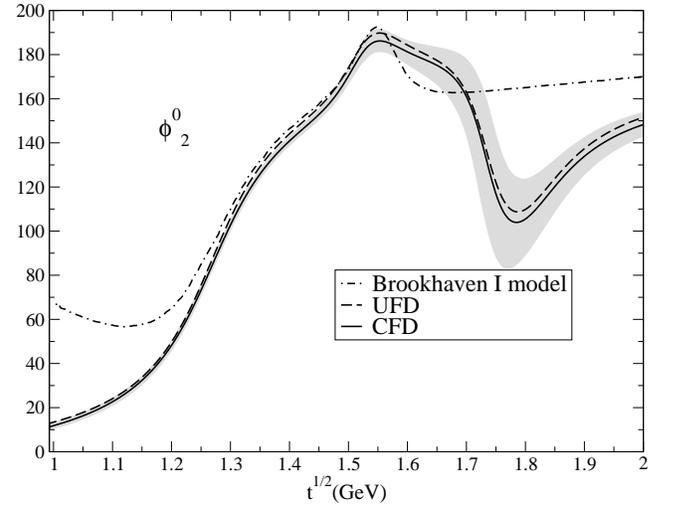}
\caption{\rm \label{fig:comparisong02cfd} Comparison between the UFD and CFD $g^0_2$ phases obtained 
with a model including an $f_2(1810)$ resonance and the one obtained with the Brookhaven model
without it, using a flat background. 
}
\end{figure}

\subsection{Constrained $g^0_2(t)$ partial wave}

For this wave the agreement was not as good as for the $I=1$ and $\ell=1$ partial wave, particularly
in the threshold region. After minimization the overall agreement has improved considerably, from $d^2=1.6 $ down to 1.1. 
However, as seen in Fig.~\ref{fig:cfd02}, our CFD parameterization still shows some small discrepancy with 
its dispersive output near threshold, although the deviation has improved substantially in that region compared to the unconstrained case. 

This improvement is achieved without changing much the CFD parameterization with respect to the UFD.
The CFD parameters change little from their previous UFD values, as seen in Table \ref{tab:g20ufd}.
In addition, in Fig.~\ref{fig:g02cfd} we can see that the deviations from the UFD to the CFD modulus are almost imperceptible. 
There are some  differences near threshold but, unfortunately, when plotting the modulus together with data, the resulting curves look almost identical due to a $q(s)^5$ factor. In contrast, we can see in Fig.~\ref{fig:comparisong02cfd}  
some small difference between the UFD and CFD phase $\phi^0_2$. This change is actually the one mostly responsible for the improvement in the $d^2$. 

We have also checked that the values obtained at the $K \bar K$ threshold still fulfill Watson's Theorem when using the $\pi \pi$ scattering values obtained from dispersion relations \cite{Bydzovsky:2016vdx, GarciaMartin:2011cn}. One should be careful not to force too much the fit in the threshold region because, as commented in the UFD case, this could spoil the $f_2(1270)$ mass, which is very well established from different experiments, not just scattering. That is why we considered the $f_2$ and $f'_2$ masses as additional data points when fitting the $\pi\pi\rightarrow K \bar K$ data. We have also added this extra contribution when minimizing the $\chi^2$ to obtain the CFD set.

We have tried different parameterizations, including additional flexibility upon Breit-Wigner-like parameterizations, but we have not been able to find a solution that satisfies better the dispersion relation near threshold without spoiling severely the data description.  

Finally, let us note that this dispersion relation has some sensitivity to $\pi K$ scattering, in particular to the scalar partial wave.  A more thorough study would require allowing the $\pi K$ scattering amplitude to vary when imposing the hyperbolic dispersion relations as constraints, but that is well beyond the scope of this work dedicated to $\pi\pi\rightarrow K \bar K$, where we have taken $\pi K$ scattering as fixed input.

\subsection{Constrained $g^0_0(t)$ partial wave}

The scalar partial wave $g^0_0$ is the most interesting in this work, given that we are dealing with
two incompatible sets of experimental data for the modulus and also because neither of them are consistent with the dispersive representation. 

As seen in Section \ref{sec:UFD}, on the one hand we have the Brookhaven-II \cite{Longacre:1986fh} data and,
on the other hand, the data of Brookhaven-I \cite{Etkin:1981sg} and Argonne \cite{Cohen:1980cq}. From these two sets we obtained the UFD$_\text{B}$ and UFD$_\text{C}$ parameterizations, respectively. For the phase we had a single UFD parameterization.
Let us recall that the overall UFD$_\text{C}$ agreement with its dispersive output up to 1.47 GeV is poor, with $d^2=2.7$, whereas the UFD$_\text{B}$ is even more inconsistent with $d^2=5.6$. In that respect the UFD$_\text{B}$ parameterization may seem disfavored.  However, the UFD$_\text{C}$ modulus is clearly incompatible with the value that would be obtained from the inelasticity of $\pi \pi$ scattering obtained from dispersion relations \cite{GarciaMartin:2011cn} assuming two coupled channels, $\pi\pi$ and $K\bar K$. For that reason we will study here both UFD$_B$ and UFD$_C$ and will obtain a fit to each data set constrained with our dispersion relation in Eq.~\eqref{MO00}. We will see that after this process both constrained solutions will be equally acceptable with respect to their consistency regarding dispersion relations.

\begin{figure}
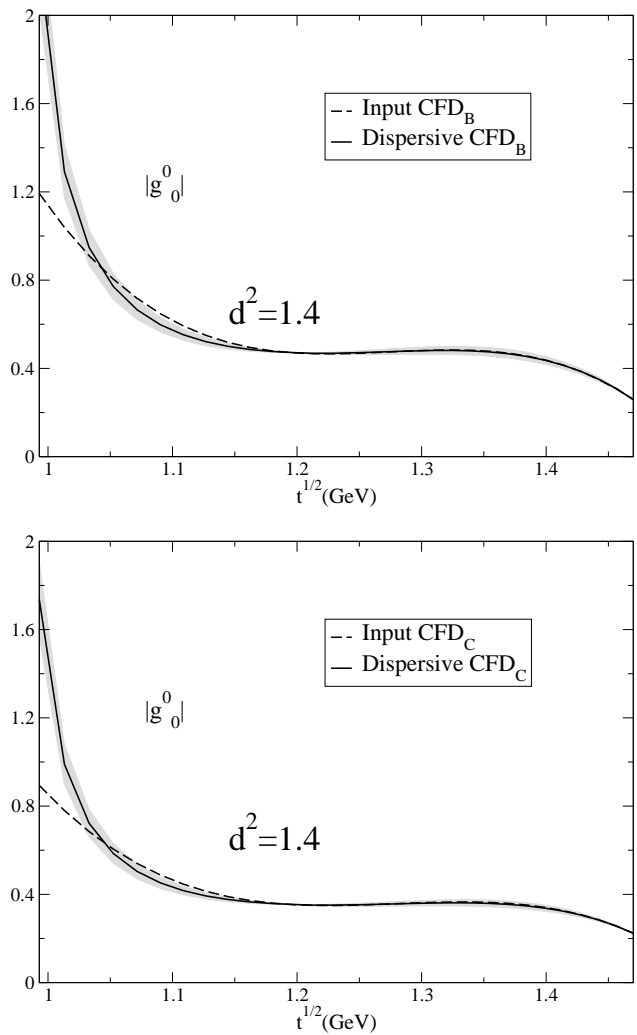

\centering
\includegraphics[scale=0.32]{checkg00cfd.eps}\vspace*{.3cm}
\includegraphics[scale=0.32]{checkg00cohencfd.eps}
\caption{Comparison between the input (dashed line) and the dispersive output (continuous line) for the modulus of the $g^0_0$ dispersion relation in Eq.\eqref{MO00}. In the upper panel we show the results using as input the CFD$_\text{B}$ parameterization and in the lower panel those from the CFD$_\text{C}$. The gray bands cover the uncertainty in the difference between the input and dispersive results. By comparing with Fig.~\ref{fig:checkg00} we see that the fulfillment of the dispersion relation by the CFD set has improved considerably with respect to the UFD parameterization. Also, there is no significant difference in the consistency of the CFD$_\text{B}$ and CFD$_\text{C}$ sets.  \label{fig:cfd00} 
}
\end{figure}

Let us note that we now use as input the $g^0_2$ CFD parameterization
obtained in the previous subsection.
The consistency test of the constrained $g^0_0$ results can be found in Fig.~\ref{fig:cfd00}. It can be seen that we obtain an equally good consistency for both
the CFD$_\text{B}$ and CFD$_\text{C}$ parameterizations except for the region very close to threshold. The behavior in this region is controlled by the $f_0(980)$ shape in the elastic region of $\pi \pi$ scattering and thus is out of the scope of this work, since we consider it input. The rest of the energy region up to 1.47 GeV has values of $d^2$ below one.

\begin{figure*}
\centering
\includegraphics[width=.9\textwidth]{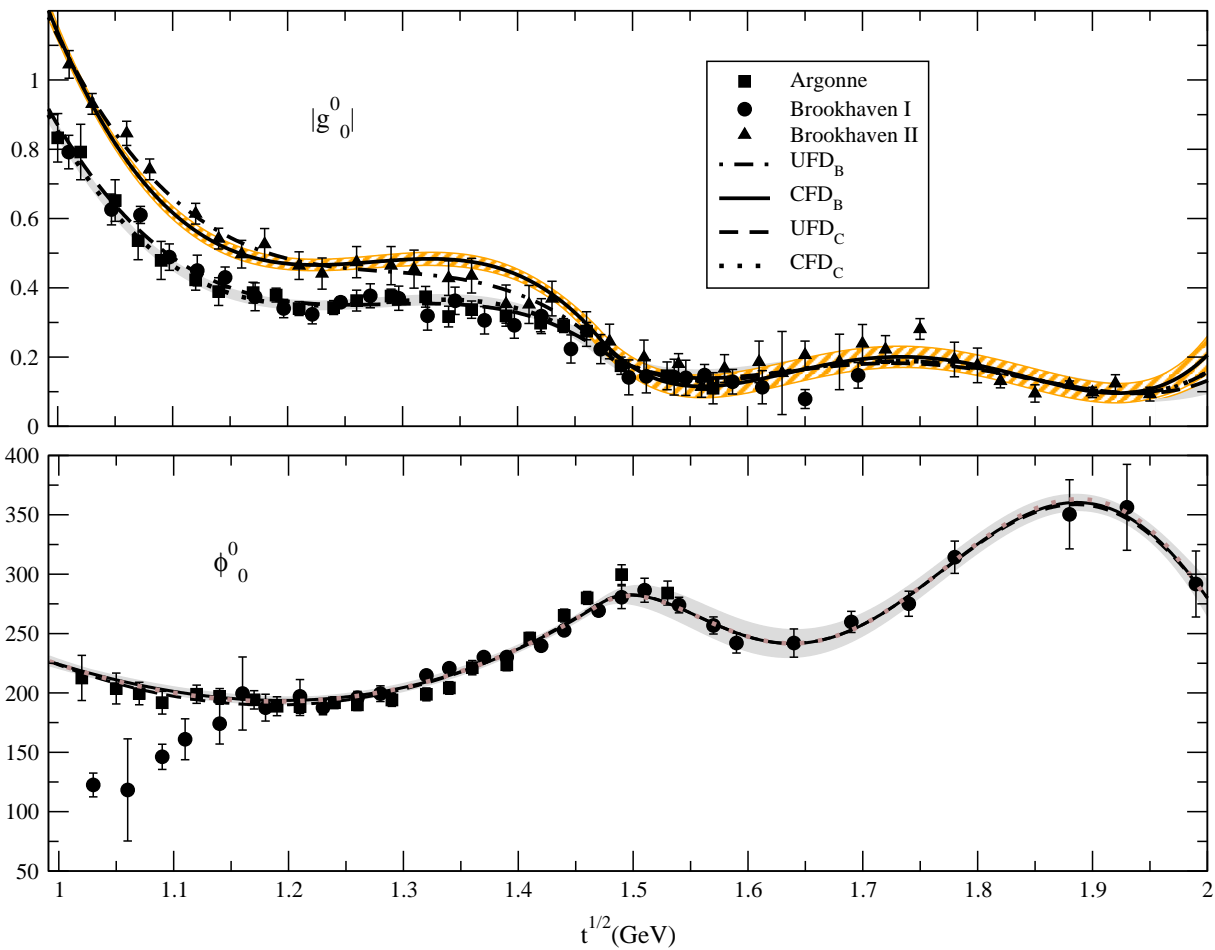}
\caption{ Comparison between the UFD and CFD parameterizations for $g_0^0(t)$.
The  bands cover the uncertainties of the CFD solutions.
Upper panel: Modulus of the scalar-isoscalar 
$\pi\pi\rightarrow K \bar K$ scattering. The dotted line represents the CFD combined fit while 
the continuous line represents the CFD fit to the Brookhaven-II data only.
The only significant change is in the 1.25 to 1.45 GeV between UFD$_\text{B}$ to CFD$_\text{B}$.
Lower panel: Scalar-isoscalar phase for $\pi\pi\rightarrow K \bar K$ scattering.
Note that the UFD, CFD$_\text{B}$ and CFD$_\text{C}$ phases are almost indistinguishable.
 \label{fig:g00cfd} }
\end{figure*}

In Fig.~\ref{fig:g00cfd} we also compare both CFD parameterizations against their respective UFD parameterizations and the data. There one can see that the UFD and CFD phases are almost identical, except in the 1.1 to 1.2 GeV region where the CFD is higher by more than one standard deviation, and in the 1.9 GeV region where the CFD phase is again higher but well within uncertainties. Actually there are two CFD$_\text{B}$ and CFD$_\text{C}$ phases but they are totally indistinguishable.

Concerning the modulus, the UFD$_\text{C}$ and CFD$_\text{C}$ are compatible, whereas the CFD$_\text{B}$ is slightly lower than
the UFD$_\text{B}$ in the 1.05 to 1.15 region, but clearly higher in the 1.3 to 1.45 region.
These differences go above the 2-$\sigma$ level, so that they lie still reasonable close to the data, but prefer to cross the top of the experimental uncertainty bars.

Note that the "dip" structure in the inelasticity from $\pi\pi$ scattering occurs around 1.1 GeV, whereas the biggest difference between the in UFD$_\text{B}$ and the CFD$_B$ is found above 1.25 GeV, so that we conclude that such a dip is not the cause of the deviation for the UFD$_\text{B}$ set. The dip structure favored by $\pi\pi$ scattering dispersive analyses can therefore be accommodated also with the hyperbolic dispersive representation of $\pi\pi\rightarrow K \bar K$.

Therefore we conclude that the data most commonly used in the literature (Argonne \cite{Cohen:1980cq}) 
is not necessarily the only acceptable solution and that one does not have to ignore the Brookhaven-II data. 
Actually, we have shown that with the CFD$_B$ solution the Brookhaven-II data can also be fairly well described 
while being consistent with $\pi\pi\rightarrow K \bar K$
dispersion relations and with the dispersive determination of the inelasticity in $\pi \pi$ 
scattering that, in contrast, is not consistent with the Argonne data. In this sense the CFD$_\text{C}$ 
is disfavored against the CFD$_\text{B}$ set.

\begin{figure}
\centering
\includegraphics[scale=0.32]{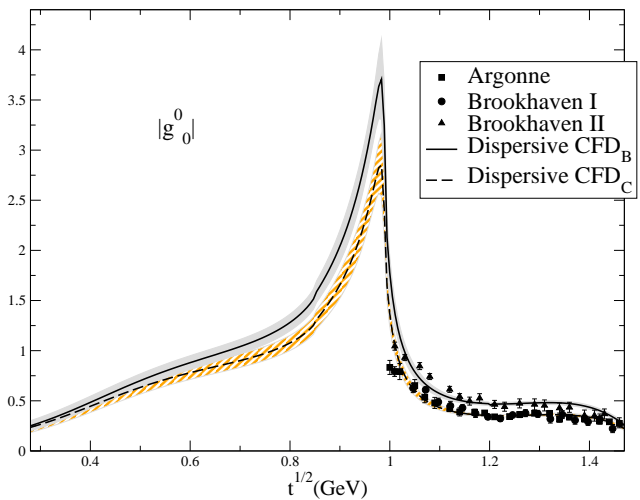}\vspace*{.3cm}
\caption{\rm \label{fig:g00cfddispersive} 
Dispersive output for the modulus of the $g_0^0(t)$ $\pi\pi\rightarrow K \bar K$ partial wave obtained from the CFD sets. Note how they differ also below the $K\bar K$ threshold.}
\end{figure}

Finally, in Fig.~\ref{fig:g00cfddispersive} we also show the CFD$_B$ and CFD$_C$ parameterizations in the unphysical region.
There  one can observe that their respective  pseudo-threshold behaviors are quite different. Namely, the modulus of the CFD$_\text{B}$ around the $f_0(980)$ peak is larger than that of the CFD$_C$. Such different behaviors may have a sizable impact for future studies of $\pi K \rightarrow \pi K$ dispersion relations.

\section{Conclusions and outlook}
\label{sec:Conclusions}

In this work we have performed a dispersive study of $\pi\pi\rightarrow K\bar K$ scattering by means of partial-wave dispersion relations of the Roy-Steiner type, i.e. based on hyperbolic dispersion relations.
While other studies with similar equations used dispersion theory to obtain information on the sub-threshold region, we have also used them  for the first time in the physical region. Moreover, we have derived a set of equations based on $(s-a)(u-a)=b$ hyperbolae in which we have obtained the value of $a$ that maximizes the applicability range of these hyperbolic dispersion relations. Compared to the existing $a=0$ case we have increased the applicability range of the hyperbolic partial-wave dispersion relations in the physical region by 67\% in the $t$ variable. This has allowed us to study dispersively the existing data sets on  $\pi\pi\rightarrow K\bar K$ up to 1.47 GeV.

In particular, on a first step we have obtained a set of unconstrained fits to data (UFD) for each partial wave $g_\ell^I(t)$, where $\ell$ and $I$ are the angular momentum and isospin, respectively. 
For the case of the scalar-isoscalar wave $g_0^0$ we have provided two alternative fits, called UFD$_\text{B}$ and UFD$_\text{C}$,
to differentiate between fits to two conflicting sets of data. In addition, we have provided high energy parameterizations for $\pi\pi\rightarrow K \bar K$ scattering, based on factorization and Regge theory, that we need for the high energy part of our dispersive integrals.
We have then tested these UFD parameterizations against our dispersion relations. We have found that the $P$ wave UFD is very consistent with dispersion relations. Also, the $D$ wave is crudely consistent with these equations, although there is clear room for improvement. In contrast, we have found that the unconstrained fits to both solutions of the scalar-isoscalar wave show a significant inconsistency with the dispersive representation, particularly, but not only, near threshold. These deviations are not related to the high energy input, and thus they become a first warning to the phenomenological use of simple fits to the existing data.

Next, we have provided a new set of fits to data using the hyperbolic partial-wave dispersion relations as constraints. For the $P$ and $D$ waves, these constrained fits to data (CFD) satisfy their dispersion relations within uncertainties while describing very well the experimental data. There is only some relatively small tension in the D-wave threshold region. In particular we have shown that a simple description of the D-wave threshold region with a simple Breit-Wigner parameterization of the nearby $f_2(1270)$ resonance is not acceptable.

We have also found that, with the exception of the region very close to threshold, both constrained parameterizations of the $g_0^0$ wave, labeled CFD$_\text{B}$ and CFD$_\text{C}$, satisfy well the dispersion relations, while still describing reasonably well their respective sets of data. Nevertheless some systematic deviations from the data central values are needed in order to satisfy the dispersive representation, particularly for the UFD$_\text{B}$ in the region between 1.25 and 1.45 GeV. This becomes a second warning towards considering only the most popular data set described by UFD$_\text{C}$: the data on which the UFD$_\text{C}$ set is based can be also described consistently with hyperbolic partial-wave dispersion relations, and is favored by previous $\pi\pi$ scattering dispersive analyses. This second set should definitely not be discarded, if not directly favored against the most popular one.

In conclusion,  our constrained data fits provide reliable, precise and simple parameterizations of data 
on S, P and D partial waves up to 2 GeV,
which are consistent with the hyperbolic dispersive representation up to its maximum applicability limit
of 1.47 GeV. 

As an outlook for this work, our constrained parameterizations could be used by both the theoretical and experimental hadron communities as input for other processes. Actually, in the near future we plan to use them for further studies. For example: to implement re-scattering effects in CP violating decays involving pions and kaons, or to study the much debated $f_0(1370)$ and $f_0(1500)$ resonance by means of model-independent methods based on analyticity, or combined with $\pi\pi$ scattering determinations, to obtain a precise determination of the  $a_0^\pm$ scattering lengths from sum rules. Finally, we will use them as input for a similar dispersive analysis of $K \pi$ scattering data and the rigorous and precise determination of light-strange resonance parameters. In particular, this input will be very useful for a precise determination of the elusive $K_0^*(800)$, by analyzing data using hyperbolic partial-wave dispersion relations of the type derived here.

\section*{Acknowledgements}

{\bf Acknowledgments} JRP and AR are supported by the Spanish project  FPA2016-75654-C2-2-P. AR would also like to acknowledge the financial support of the Universidad Complutense de Madrid through a predoctoral scholarship.  We would also like to thank B. Moussallam and J. Ruiz de Elvira for fruitful discussions, as well as J. Miranda for her comments and corrections.

\appendix

\section{Modified $g^0_0(t)$ data extraction above 1.6 GeV}
\label{sec:modifiedg00}

In the main text we have included a third pole for the $f_2(1810)$ in the $g^0_2(t)$ partial wave, since it is listed in the RPP, although it claims that "Needs confirmation".
As we already commented, this produces a large oscillation of the phase above 1.6 GeV, different from the almost flat parameterization used in \cite{Etkin:1981sg}, as can be seen in Fig. \ref{fig:comparisong02}. 

\begin{figure}[h]
\includegraphics[scale=0.32]{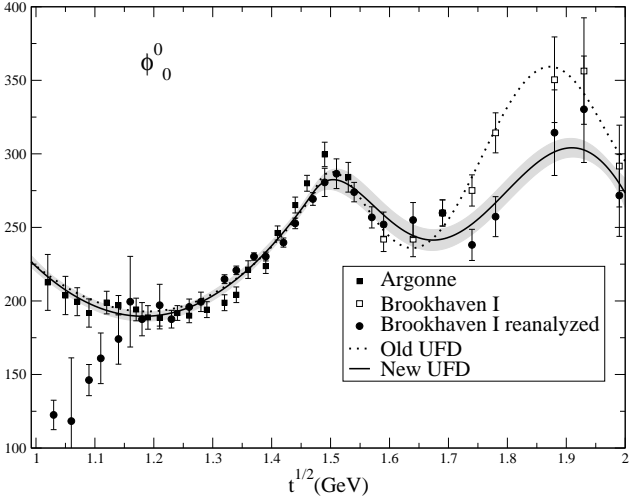}\vspace*{.3cm}
\caption{\rm \label{fig:g00newphase} 
New $\phi^0_0$ phase obtained after extracting the data from \cite{Etkin:1981sg}  by using our UFD for the $g^0_2$ partial wave
}
\end{figure}

However, in \cite{Etkin:1981sg} the $g^0_2(t)$ wave is used as input to extract the $g^0_0(t)$.
Hence, if one now assumes the existence of the $f_2(1810)$, the extraction of the $g^0_0(t)$ phase above 1.6 GeV no longer corresponds to the one given in the paper. The "New UFD" $g^0_0(t)$ phase we obtain is shown in Fig. \ref{fig:g00newphase}, which parameters can be found in Table \ref{tab:g00newphase}.  Let us recall that above 1.6 GeV the modulus is rather small, so that its contribution to the dispersion relation below 1.47 GeV is also very small. However, one may still wonder if this new UFD S-wave  phase above 1.6 GeV could change significantly the results for the modulus after analyzing the dispersion relations.

\begin{table}[h] 
\caption{Parameters of the new $\phi^0_0$.} 
\centering 
\begin{tabular}{c c c c} 
\hline\hline  
\rule[-0.05cm]{0cm}{.35cm}Parameter & New UFD & New CFD$_\text{B}$ & New CFD$_\text{C}$ \\ 
\hline 
\rule[-0.05cm]{0cm}{.35cm}$B_1$ & 23.5  $\pm$1.3  & 21.8  $\pm$1.3 & 22.5  $\pm$1.3\\ 
\rule[-0.05cm]{0cm}{.35cm}$B_2$ & 29.0  $\pm$1.3  & 27.3  $\pm$1.3 & 27.9  $\pm$1.3\\ 
\rule[-0.05cm]{0cm}{.35cm}$B_3$ & 0.01  $\pm$1.60  & 1.49  $\pm$1.60 &   0.81  $\pm$1.60\\ 
\rule[-0.05cm]{0cm}{.35cm}$C_1$ & 12.0890 fixed  & 12.4388 fixed & 12.1076 fixed\\ 
\rule[-0.05cm]{0cm}{.35cm}$C_2$ & 13.6  $\pm$2.6  & 13.6  $\pm$2.6 & 13.3  $\pm$2.6\\ 
\rule[-0.05cm]{0cm}{.35cm}$C_3$ &-12.9 $\pm$2.3  &-13.0 $\pm$2.3 &-13.1 $\pm$2.3\\ 
\rule[-0.05cm]{0cm}{.35cm}$C_4$ &-13.1 $\pm$2.2  &-13.3 $\pm$2.2 &-13.4 $\pm$2.2\\ 
\rule[-0.05cm]{0cm}{.35cm}$C_5$ &  4.0 $\pm$2.4  &  4.2 $\pm$2.4 &  3.9 $\pm$4.0\\ 
\hline 
\end{tabular} 
\label{tab:g00newphase} 
\end{table}

\begin{figure}\vspace*{.3cm}
\includegraphics[scale=0.32]{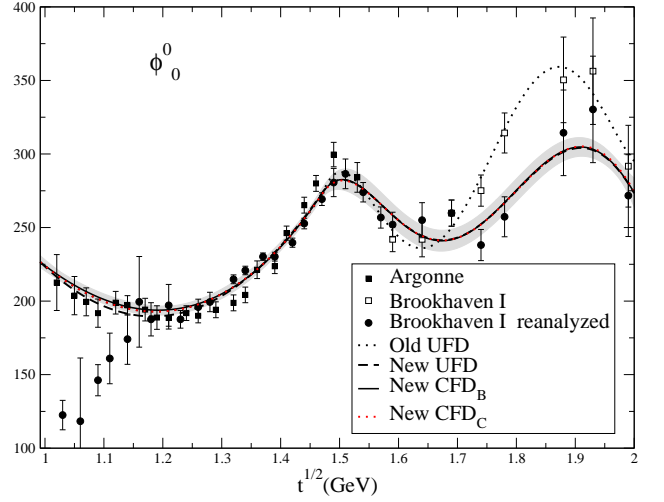}
\caption{\rm \label{fig:g00newphasecfd} 
New CFD $\phi^0_0$ phase obtained after extracting the data from \cite{Etkin:1981sg} by using our model for the $g^0_2$ partial wave
}
\end{figure}

Hence, we have run again our whole procedure to obtain a "New CFD" phase for $g^0_0(t)$ and we show in Fig. \ref{fig:g00newphasecfd} the final result of the new analysis. As expected, since the input is small above 1.6 GeV, the values obtained for the modulus  are almost equal to the ones calculated with the old phase and we do not plot them.

However, as a matter of fact, the $g^0_0(t)$ phase above 1.6 GeV is different if one assumes the presence of the $f_2(1810)$ in the $g^0_2(t)$. If one wants to be consistent with that assumption, which at present in the RPP seems to be favored  versus the flat solution used in \cite{Etkin:1981sg}, then one should use our "New UFD" rather than the main one in the text. Of course, the difference below 1.47 GeV is negligible.

\section{Kernels}
\label{app:kernels}

In this section we provide the explicit expressions for the $G^I_{\ell \ell'}(t,t')$ and $G^\pm_{\ell \ell'}(t,s')$ kernels needed in the partial-wave dispersion relations in Eq.~\eqref{eq:pwhdr}. Recall that $\ell\leq2$ corresponds to the angular momentum of the partial-wave dispersion relation, i.e. the ``output" partial wave, whereas $\ell'$ corresponds to the angular momentum
of the ``input" wave in the integrand of the dispersion relation. 
Similarly, $s'$ and $t'$ are the integration variables, whereas $t$ is the variable of the ``output" partial wave
coming out of the dispersion relation. 
Note that,
in the input, partial waves with $\ell'>2$ can be safely neglected, except for the $\ell'=4$ partial wave needed for the $g^0_2$ equation, which nevertheless gives a rather small contribution.

Let us first recall some previous definitions:
\begin{eqnarray}
z_{s'}&=&1+\frac{2s't}{\lambda_{s'}}, \nonumber \\
\lambda_{s'}&=&(s'-(m_\pi+m_K)^2)(s'-(m_\pi-m_K)^2)\,.\nonumber 
\label{anglet}
\end{eqnarray}

We start by listing the kernels of the $g^1_1(t)$ partial wave:
\begin{widetext}
\begin{eqnarray}
G^1_{1,3}(t,t')&\!\!=&\frac{7}{48}(t+t'-4\Sigma+10a), 
\label{kernels11} \\
G^-_{1,0}(t,s')&\!\!\!=&\!\!\!4\sqrt{2}\left[
\frac{(2s'-2\Sigma+t)A(t,s')-4q_K(t)q_{\pi}(t)}{16(q_K(t)q_{\pi}(t))^3} \right], \nonumber \\
G^-_{1,1}(t,s')&\!\!\!=&\!\!\!12\sqrt{2}\left[P_1(z_{s'})
\frac{(2s'-2\Sigma+t)A(t,s')-4q_K(t)q_{\pi}(t)}{16(q_K(t)q_{\pi}(t))^3}-\frac{2s'}{3(s'-a)\lambda_{s'}} \right], \nonumber \\
G^-_{1,2}(t,s')&\!\!\!=&\!\!\!20\sqrt{2} \left[P_2(z_{s'})
\frac{(2s'-2\Sigma+t)A(t,s')-4q_K(t)q_{\pi}(t)}{16(q_K(t)q_{\pi}(t))^3}-\frac{2s'z_s'}{(s'-a)\lambda_{s'}}+\frac{s'^2(2s'+t-2\Sigma)^2}{2(s'-a)^2\lambda_{s'}^2}
-\frac{24s'^2(q_K(t)q_\pi(t))^2}{5(s'-a)^2\lambda_s'^2} \right],\nonumber
\end{eqnarray}
\end{widetext}
where $P_l(z_{s'})$ are the Legendre polynomials, $a$ is one of the parameters that defines the hyperbola $(s-a)(u-a)=b$ and we have defined 
$$A(t,s')={\rm Arcth}\Big(\frac{4q_K(t)q_{\pi}(t)}{2s'+t-2\Sigma}\Big).$$

For the $g^0_2$ case we first define for convenience $$x(t,s')=\frac{4q_K(t)q_{\pi}(t)}{2s'+t-2\Sigma}.$$ 

\begin{widetext}
By using the same definitions as above one obtains 

\begin{eqnarray}
G^0_{2,4}(t,t')&\!\!=&\!\!\frac{3}{8} (t+t'-4\Sigma+7a), 
	\nonumber \\
G^+_{2,0}(t,s')&\!\!=&\!\!\frac{\sqrt{3}(2s'+t-2\Sigma)^2}{32  q_K(t)^5
   q_\pi(t)^5}
\left[(3-x(t,s')^2) A(t,s')-3x(t,s')\right], \nonumber \\
G^+_{2,1}(t,s')&\!\!=&\!\!\frac{3\sqrt{3}(2s'+t-2\Sigma)^2}{32 q_K(t)^5
   q_\pi(t)^5} 
P_1(z_{s'}) \left[(3-x(t,s')^2) A(t,s')-3x(t,s')\right] , \nonumber \\
G^+_{2,2}(t,s')&\!\!=&\!\!5\sqrt{3}
\left[\frac{(2s'+t-2\Sigma)^2}{32 q_K(t)^5
   q_\pi(t)^5 }P_2(z_{s'})\Big((3-x(t,s')^2) A(t,s')-3x(t,s')\Big)-\frac{16s'^2t}{5(s'-a)^2\lambda_{s'}^2}\right]. 
	\label{kernels02}
\end{eqnarray}
Finally, for the $g_0^0(t)$ dispersion relation the kernels we need are
\begin{eqnarray}
G^0_{0,2}(t,t')&\!\!=&\!\!\frac{5}{16}(t+t'-4\Sigma+6a), \nonumber	 \\
G^+_{0,0}(t,s')&\!\!=&\!\!\sqrt{3}\left[\frac{A(t,s')}{q_K(t)q_{\pi}(t)}  +\frac{2(\Sigma-s')}{\lambda_{s'}}\right], \nonumber \\
G^+_{0,1}(t,s')&\!\!=&\!\!3\sqrt{3}\left[\frac{A(t,s')}{q_K(t)q_{\pi}(t)}P_1(z_{s'})
-\frac{(2s'+2t-2\Sigma)}{\lambda_{s'}}-\frac{2at}{(s'-a)\lambda_{s'}}\right], \nonumber \\
G^+_{0,2}(t,s')&\!\!=&\!\!5\sqrt{3}\left[\frac{A(t,s')}{q_K(t)q_{\pi}(t)}
P_2(z_{s'})
-\frac{2s-2\Sigma}{\lambda_{s'}}-\frac{6st(\Delta^2+s'(3s'+2t-4\Sigma)}{(s'-a)\lambda_{s'}^2}+\frac{3s'^2t(2s'+t-2\Sigma)^2}{2(s'-a)^2\lambda_s'^2}\right.\nonumber\\
&&\quad\left.-\frac{8s'^2t(q_K(t)q_{\pi}(t))^2}{(s'-a)^2\lambda_{s'}^2}\right].
\label{kernels00}
\end{eqnarray}

All these kernels produce smooth integrable inputs in the physical region. They also produce the left and circular cut structures required by partial wave projection.

\vspace{0.5 cm}

\end{widetext}

\section{$t$-channel numerical solution}

In order to calculate numerically the Omn\`es integrals it is convenient
to make a change of variables to facilitate the integration near $t_m$. 
For concreteness we explain the $g^1_1(t)$ dispersion relation, 
following closely the method explained in \cite{Buettiker:2003pp,Hoferichter:2011wk}
although in our case it has one less subtraction.
The other waves are similar.
We start by separating within the integrals the regions above and below $t_m$, 
\begin{align}
&g^1_1(t)=\Delta^1_1(t)+\frac{\Omega^1_1(t)}{\pi}\left[\rule[-0.05cm]{0.cm}{.5cm}\right.\\
&\quad \int^{t_m-\tau}_{4m_\pi^2}\!\!\!\!\!dt'\frac{\Delta^1_1(t')\sin \phi^1_1(t')}{\Omega^1_{1,R}(t')(t'-t)} +\int^{t_m}_{t_m-\tau}\!\!\!\!\!dt'\frac{\Delta^1_1(t')\sin \phi^1_1(t')}{\Omega^1_{1,R}(t')(t'-t)} \nonumber\\
&+\left.\int^{\infty}_{t_m+\tau}\!\!\!\!\!dt' \frac{\vert g^1_1(t')\vert \sin \phi^1_1(t')}{\Omega^1_{1,R}(t')(t'-t)} +\int^{t_m+\tau}_{t_m}\!\!\!\!\!dt' \frac{\vert g^1_1(t')\vert \sin \phi^1_1(t')}{\Omega^1_{1,R}(t')(t'-t)}\right].
\nonumber
\end{align} 

We now introduce the variable $v(t')=(t'-t_m)/(t_m-t)$ and write:
\begin{align}
&\frac{\Omega(t)}{\pi}\int^{t_m}_{t_m-\tau}\!\!\!\!\!dt'\frac{\Delta^1_1(t')\sin \phi^1_1(t')}{\Omega^1_{1,R}(t')(t'-t)}= \nonumber \\
&\quad\frac{\Delta^1_1(t_m)\exp(i \phi^1_1(t_m))\sin \phi^1_1(t_m)}{\pi}\int_0^{\tau(t)}\!\!\!\!\!\frac{dv}{v^{\phi^1_1(t_m)/\pi}(1-v)}, \nonumber \\
&\frac{\Omega(t)}{\pi}\int^{t_m+\tau}_{t_m}\!\!\!\!\!dt' \frac{\vert g^1_1(t')\vert \sin \phi^1_1(t')}{\Omega^1_{1,R}(t')(t'-t)}= \nonumber \\
&\qquad\qquad\frac{g^1_1(t_m)\sin \phi^1_1(t_m)}{\pi}\int^{\tau(t)}_0\!\!\!\!\!\frac{dv}{v^{\phi^1_1(t_m)/\pi}(1+v)}.
\end{align}

As shown in \cite{Buettiker:2003pp} this equation also implies the continuity of the partial waves at the matching point $t_m$.  Since $\tau(t_m)=\infty$ and using
\begin{align}
&\frac{1}{\pi}\int_0^{\infty}\frac{dv}{v^{\phi^1_1(t_m)/\pi}(1-v)}=-\frac{\exp(-i \phi^1_1(t_m))}{\sin(\phi^1_1(t_m))}, \nonumber \\
&\frac{1}{\pi}\int^{\infty}_0\frac{dv}{v^{\phi^1_1(t_m)/\pi}(1+v)}=\frac{1}{\sin(\phi^1_1(t_m))},
\end{align}
inside Eqs.\eqref{MO00},\eqref{MO11},\eqref{MO20} one recovers the matching values $\vert g^0_0(t_m) \vert$, $\vert g^1_1(t_m) \vert$, $\vert g^0_2(t_m) \vert$. In addition, for $g^0_0$, and due to the introduction of the free parameter $\alpha$, one has to impose  a smooth continuity condition at $t_m$ to fix $\alpha$, which is done numerically in this work. Otherwise spurious cusps would be produced for the modulus of the amplitude at $t=t_m$, spoiling the analytic structure and its behavior at different values of $t$.

\section{Applicability Range}
\label{app:convergence}

Let us recall that in this work our aim is to maximize the applicability range of
the partial-wave hyperbolic dispersion relations in the real axis, by choosing the $a$ parameter appropriately. 
Our approach will be similar to that in \cite{Ditsche:2012fv, Hoferichter:2011wk} and we will study the applicability range both for the $s$-channel $\pi K \rightarrow \pi K$ and for the $t$-channel $\pi \pi \rightarrow K \bar K$. 

\begin{figure}
\centering
\centerline{ \includegraphics[width=0.45\linewidth]{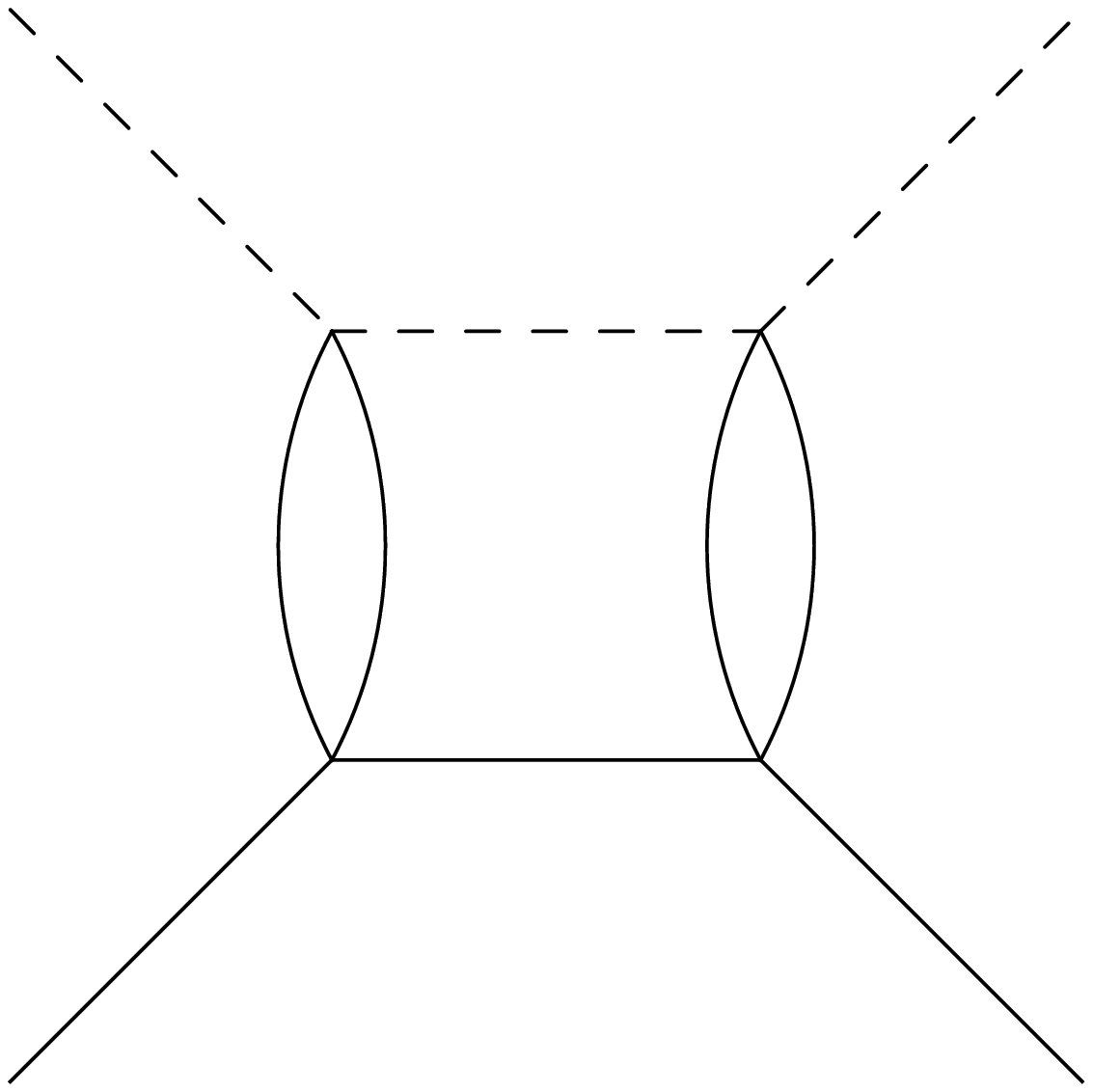} \hspace{0.1cm} \includegraphics[width=0.45\linewidth]{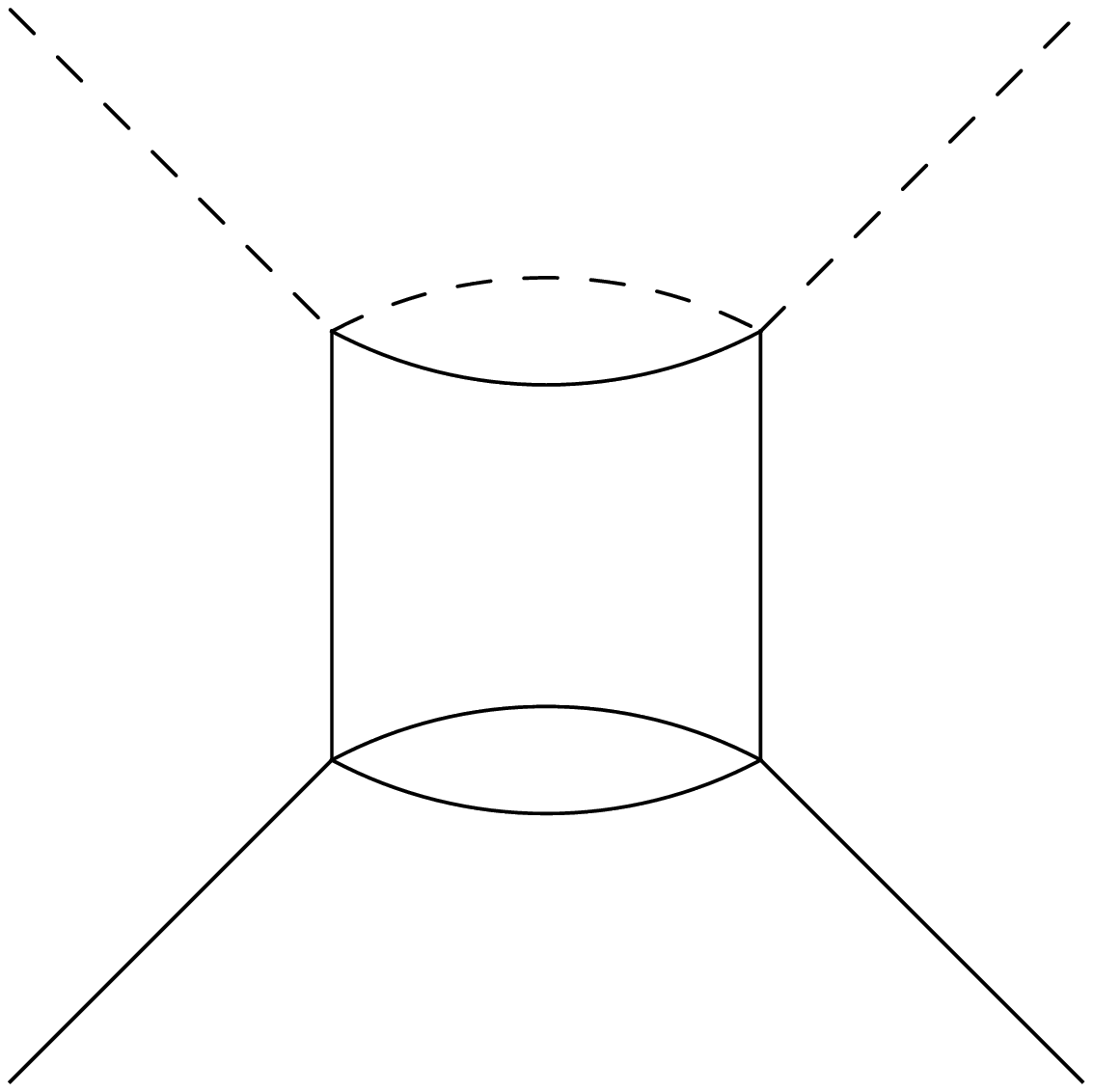}}
\vspace{0.2cm}
\centerline{ \includegraphics[width=0.45\linewidth]{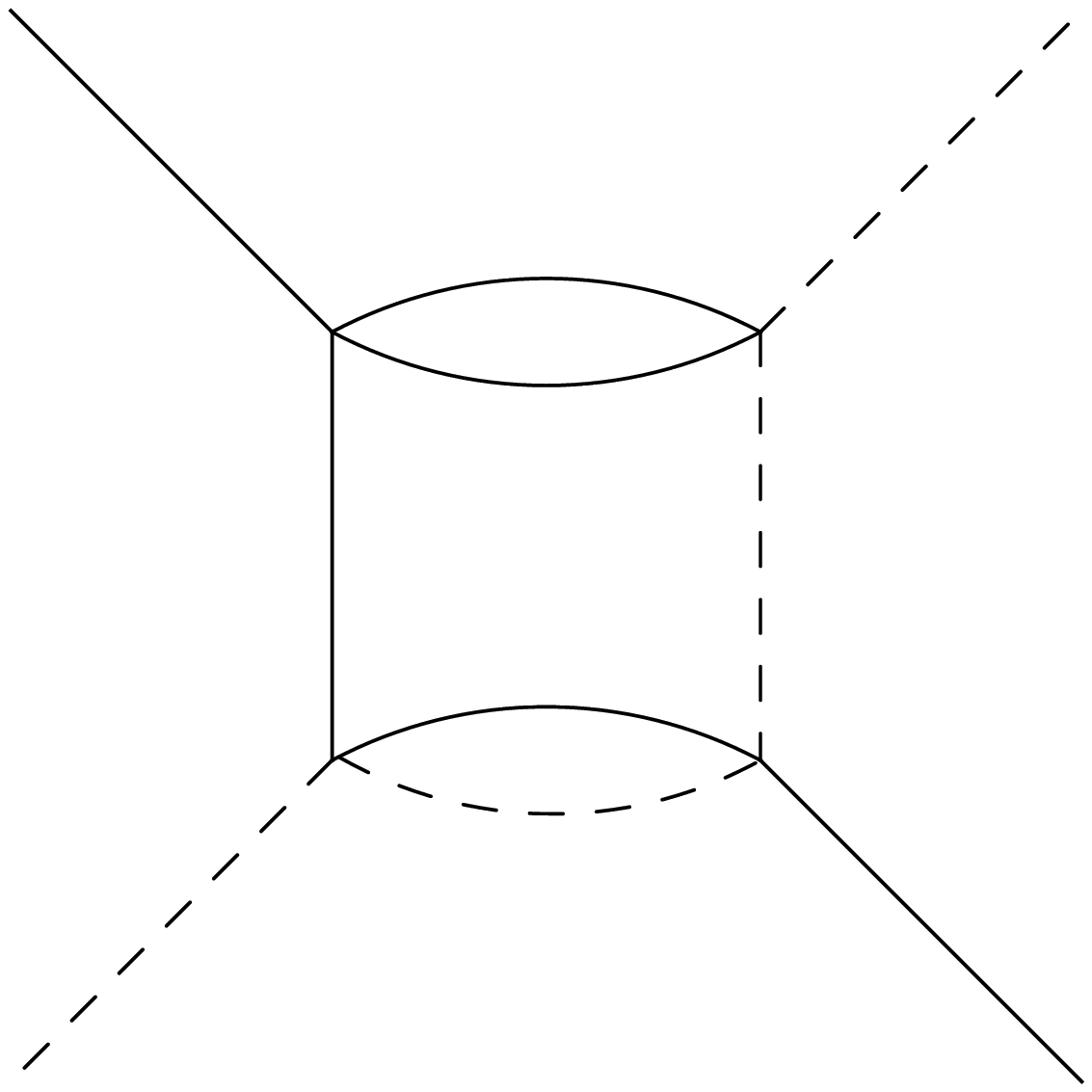} \hspace{0.1cm} \includegraphics[width=0.45\linewidth]{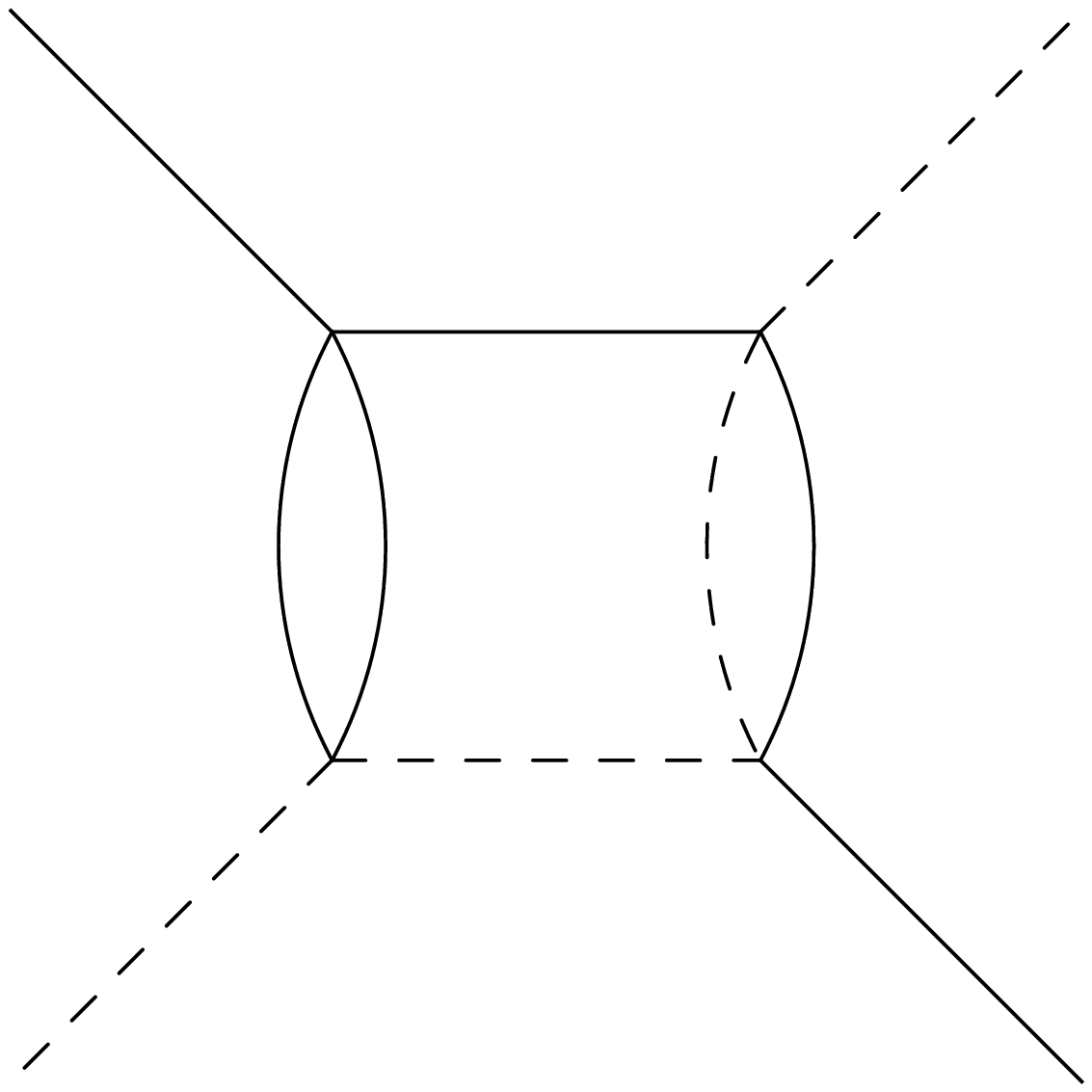}}
\caption{Box diagrams for $\pi K$ scattering. Continuous lines denote pions while dashed lines denote kaons. \label{fig:boxdiabrams}}
\end{figure}

First of all we have to calculate the double spectral regions, where the imaginary part of the amplitude becomes also imaginary and therefore the Mandelstam hypothesis does not hold (see \cite{MartinSpearman} for a textbook introduction).
For this we use the $\pi K$ scattering box diagrams that we show in Fig.~\ref{fig:boxdiabrams} (see also
\cite{DescotesGenon:2006uk}). Then we obtain
the restrictions needed to avoid these regions when projecting into partial waves 
for all the $s, t$ and $u$ channels. 
In addition, one has to ensure that the partial-wave projection is used only inside the so-called Lehmann ellipse
\cite{Lehman-Martin},  where its convergence is guaranteed.
Finally by considering the strongest 
restriction we maximize the domain of applicability by fixing $a$.

\subsection{Double spectral regions}

The equations that describe the boundary of the support of the spectral function $\rho_{s t}$ are:
\begin{align}
&b_I(s,t): (t-16m_\pi^2)\lambda_s-64m_\pi^4 s=0, \label{eq:boundaryst} \\
&b_{II}(s,t):(t-4m_\pi^2)(s-(m_K+3m_\pi)^2)-32m_\pi^3m_+=0.\nonumber
\end{align}
By means of $s \leftrightarrow u$ crossing, similar equations are obtained for $\rho_{u t}$. The equations that describe the boundary of the support of $\rho_{u s}$ are
\begin{align}
&b_{III}(s,u):\\
&\quad (s-(m_K-m_\pi)^2)(t+s-(m_K+m_\pi)^2) \nonumber \\
&\times ((m_K^2+2 m_K m_\pi+5 m_\pi^2-s)^2 \nonumber \\
&+t (s-(m_K+3
   m_\pi)^2))
=0, \nonumber \\
&b_{IV}(s,u):\nonumber\\
&\quad (s-(m_K - m_\pi)^2)(t+s-(m_K + m_\pi)^2) \nonumber \\
&\times ((( 3 m_\pi-m_K) (m_K + m_\pi) + s)^2 \nonumber \\
&+t(s-(m_K + m_\pi)^2))=0,
\end{align}
where $\lambda(x,y,z)=x^2+y^2+z^2-2xy-2xz-2yz$.

Out of these three possible spectral regions, the most restrictive boundary is that of the $\rho_{s t}$ support. Thus, by solving Eq.~\eqref{eq:boundaryst} for $t$ as a function of $s$ one obtains
\begin{align}
&T_{st}(s)=16m_\pi^2+\frac{64m_\pi^4s}{\lambda_s}, \hspace{0.3cm}\forall s\leq s_0,\\
&T_{st}(s)=4m_\pi^2+\frac{32m_\pi^3(m_K+m_\pi)}{(s-(m_K+3m_\pi)^2)},  \hspace{0.3cm}\forall s\geq s_0,
\label{eq:boundaryfort}
\end{align}
where
\begin{equation}
s_0=m_K^2+4m_Km_\pi+5m_\pi^2+2m_\pi\sqrt{5m_K^2+12m_Km_\pi+8m_\pi^2}.
\end{equation}

As shown in \cite{Steiner:1971ms}, the most simple set of curves in the Mandelstam plane that combine both crossed channels, do not introduce complicated kernels and are suitable to study partial waves in a wide range, are hyperbolas defined trough the relation $(s-a)(u-a)=b$.


In the next subsection we will combine the double spectral region constraints with those restrictions arising from the partial wave projection.

\subsection{Lehmann ellipse}

We now have to consider the projection of $T(s,t,u)$ into partial waves for the two different channels that appear in the hyperbolic dispersion relations. 

Thus, on the one hand, for a fixed value of $a$, the family of hyperbolas $(s-a)(u-a)= b$ must not enter any double spectral region for all values of $b$ needed to perform the partial-wave projection. On the other hand, for a fixed $a$, we now calculate the restriction on $b$ implied by requiring to stay within the Lehmann ellipse. This depends on what channel we perform the partial-wave projection.

\subsubsection{$s$-channel}

The partial-wave expansion for the $s$-channel converges for angles $z_{s'}(s',t')=1+2s't'/\lambda_{s'}$ inside the Lehmann ellipse \cite{Lehman-Martin} 
\begin{equation}
\frac{(\re z_{s'})^2}{A^2_s}+\frac{(\im z_{s'})^2}{B^2_s}=1,
\end{equation}
where the foci are located at $z_{s'}=\pm1$. The maximum value of $z_{s'}$ that does not enter inside the double spectral region is obtained for $t'=T_{st}(s')$, namely
\begin{equation}
z^{max}_{s'}=1+\frac{2s'T_{st}(s')}{\lambda_{s'}}=A_s,
\quad \forall s'\geq m_+^2,
\end{equation}
with the constraint given by the ellipse
\begin{equation}
-z^{max}_{s'}\leq z_{s'}\leq z^{max}_{s'}.
\end{equation}

This relation translates into a restriction on $t'$
\begin{equation}
-\frac{\lambda_{s'}}{s'}-T_{st}(s')\leq t'\leq T_{st}(s').
\end{equation}
Now, by using $b(s,t,a)=(s-a)(2\Sigma-s-t-a)$ we obtain the following set of bounds for $b$:
\begin{align}
&b^-_s(s',a)\leq b\leq b^+_s(s',a), \nonumber \\
&b^-_s(s',a)=(s'-a)(2\Sigma-s'-T_{st}(s')-a),  \nonumber \\
&b^+_s(s',a)=(s'-a)(2\Sigma-s'+\frac{\lambda_{s'}}{s'}+T_{st}(s')-a)  .
\end{align}

Thus, the final range of values allowed for $b$ to avoid touching any boundary are
\begin{equation}
b^-_s(a)\leq b\leq b^+_s(a),
\end{equation}
where
\begin{align}
&b^-_s(a)=\min b^-_s(s',a),\nonumber \\
&b^+_s(a)=\max b^+_s(s',a).
\end{align}

\subsubsection{$t$-channel}

The argument is now more complicated due to the non-linear relation between the scattering angle and $t'$ for the $t$-channel partial wave projection
\begin{equation}
z^2_{t'}=\frac{(t'-2\Sigma+2a)^2-4b(s',t',a)}{16q_\pi(t')^2 q_K(t')^2},
\end{equation}
so we use the ellipse for $z_{t'}^2$
\begin{equation}
\frac{(\re z^2_{t'}-\frac{1}{2})^2}{\hat{A}^2_t}+\frac{(\im z^2_{t'})^2}{\hat{B}^2_t}=1,
\end{equation}
where $\hat{A}_t=(A^2_t+B^2_t)/2$ and $\hat{B}_t=A_tB_t$ are the axes of the ellipse for $z^2_{t'}$ and $A_t,B_t$ the ones for $z_{t'}$. Then, the geometrical restrictions for $z^2_{t'}$ are
\begin{align}
&1-A^2_t\leq z^2_{t'}\leq A^2_t.
\label{eq:georestriction}
\end{align}

As shown in Eq.~\eqref{eq:anglet} the relation between $z_t$ and $s-u$ is really simple, calling $\nu=s-u$ and rewriting equation \eqref{eq:boundaryfort} in terms of $\nu$ we obtain
\begin{align}
&\nu_{st}(t)=\frac{-16m_\pi^3m_K-12m_\pi m_+t-t^2}{4m_\pi^2-t},   \hspace{0.3cm}\forall t\geq t_\pi, \nonumber \\
&\nu_{st}(t)=\frac{1}{t-16m_\pi^2} \nonumber\\
&\qquad\times\Big[ (t-8m_\pi^2)^2+4m_\pi \sqrt{t}\sqrt{(t-16m_\pi^2)m_K^2+16m_\pi^4)}  \Big] ,\nonumber\\ 
&\hspace{2cm}\forall t\geq 4t_\pi,
\end{align}
Defining now the upper bound as
\begin{equation}
N_{st}(t)=\min \nu_{st}(t),
\label{eq:bound}
\end{equation}
we obtain that 
\begin{equation}
z^{\max}_{t'}(t')=\frac{N_{st}(t')}{4q_\pi(t') q_K(t')}=A_t \hspace{0.3cm}\forall t'\geq t_K,
\end{equation}
now using equation \eqref{eq:georestriction} together with \eqref{eq:bound} we obtain the restriction for $\nu$

\begin{equation}
16[q_\pi(t') q_K(t')]^2-N_{st}(t')^2\leq \nu^2\leq N_{st}(t')^2,
\end{equation}
finally, the restriction for $b$ is obtained just by translating the $\nu^2=(t'-2\Sigma+2a)^2-4b$ constraint 
into
\begin{equation}
b^-_t(t',a)\leq b\leq b^+_t(t',a),
\end{equation}
with
\begin{align}
&b^-_t(t',a)=\frac{(t'-2\Sigma+2a)^2-N_{st}(t')^2}{4}, \nonumber \\  
&b^+_t(t',a)=\frac{(t'-2\Sigma+2a)^2-16(q_\pi(t') q_K(t'))^2+N_{st}(t')}{4}.
\end{align}

Defining again the bounds
\begin{align}
&b^-_t(a)=\max b^-_t(t',a), \nonumber \\
&b^+_t(a)=\min b^+_t(t',a),
\end{align}
we have finally obtained the allowed values of $b$ for a fixed $a$ that do not touch any boundary while projecting $t$-channel partial waves
\begin{equation}
b^-_t(a)\leq b\leq b^+_t(a),\hspace{0.3cm}\forall t\geq t_\pi\geq a.
\end{equation}

\subsection{Partial-wave Projection}

\subsubsection{$s$-channel}

Hence, to perform the partial-wave projection for the $s$-channel 
we must require $b\in[b^-_{s,t}(a),b^+_{s,t}(a)]$. For this to occur, 
we need $s\leq s_{max}$, where $s_{max}$ is the value of $s$
for which the region of projection touches the support of the double spectral region.
Since the integration range $-1\leq z_s\leq 1$ translates into
\begin{equation}
-\frac{\lambda_s}{s}\leq t\leq 0,
\end{equation}
then, given a fixed $a$, the limits on $b$ due to the $s$-channel projection are
\begin{align}
&b^{min}(s,a)\leq b\leq b^{max}(s,a), \nonumber \\
&b^{min}(s,a)=(s-a)(2\Sigma-s-a), \nonumber \\
&b^{max}(s,a)=(s-a)(2\Sigma-s+\frac{\lambda_s}{s}-a).
\end{align}

Now, $s_{max}$ is reached when touching the Lehmann ellipse, namely
\begin{align}
&b^{min}(s_{max},a)=b^-_{s,t}(a), \nonumber \\
&b^{max}(s_{max},a)=b^+_{s,t}(a).
\end{align}

We can now choose $a$ to obtain the largest $s_{max}$ and thus maximize the projection region.
For the $s$-channel projection the strongest restriction comes from the $t$-channel Lehmann ellipse
and therefore
\begin{align}
&a=-13.9 \,m_\pi^2, \hspace{0.5 cm} s_{max}=0.98 \,\gev^2 , \nonumber \\
&b^-_t(a)=-592\, m_\pi^4, \hspace{0.5 cm} b^+_t(a)=1070 \,m_\pi^4.
\end{align}

\subsubsection{$t$-channel}

To perform the $t$-channel projection we need to consider the scattering angle
\begin{equation}
0\leq z^2_t=\frac{(t-2\Sigma+2a)^2-4b}{16q^2_\pi q^2_K}\leq 1.
\label{proyection}
\end{equation}
\vspace{0.2 cm}
To maximize the domain using $a$ we search for the value $t=t_{max}$ where both the maximum and minimum values of $b$ coincide with $b^-_{s,t}(a)$ and $b^+_{s,t}(a)$. Using Eq.\eqref{proyection} and taking into account that the projection is made between $z^2_t=0$ and $z^2_t=1$ this means
\begin{align}
&z^2_t(t_{max},b^-_{s,t}(a))=1, \nonumber \\
&z^2_t(t_{max},b^+_{s,t}(a))=0.
\end{align}

Once again, the restriction of the $t$-channel is stronger than the one of the $s$-channel, and therefore
\begin{align}
&a=-10.9 m_\pi^2, \hspace{0.5cm}  -0.286 \, \gev^2\leq t\leq 2.19\, \gev^2, \nonumber \\
&b^-_t(a)=-672 \, m_\pi^4, \hspace{0.5cm} b^+_t(a)=1010 \, m_\pi^4.
\end{align}

Note that the upper limit for $t\simeq \sqrt{2.19}\,\gev\simeq1.47\,\gev$, which is the value we have been using throughout this work as the maximum applicability range of our partial-wave hyperbolic dispersion relations.
Taking these values into account one can proceed to study the physical region of both processes. 
 Note that HDR are a very useful tool to study the crossed channel and extend as much as possible
the applicability range in its real axis. However their convergence in the real axis of the $s$-channel is worse than for fixed-$t$ dispersion relations. Nevertheless, the scope of this work is precisely the study of the $t$-channel partial waves, and therefore HDR are best suited for our purposes.

\end{document}